\documentclass{mn2e}
\usepackage{graphicx}

\def\go{
\mathrel{\raise.3ex\hbox{$>$}\mkern-14mu\lower0.6ex\hbox{$\sim$}}
}
\def\lo{
\mathrel{\raise.3ex\hbox{$<$}\mkern-14mu\lower0.6ex\hbox{$\sim$}}
}
\def\simeq{
\mathrel{\raise.3ex\hbox{$\sim$}\mkern-14mu\lower0.4ex\hbox{$-$}}
}

\def\etal{et al.\ }

\def\msun{{\rm M_{\odot}}}

\begin{document}

\title[{\it XMM-Newton} observations of seven soft X-ray excess QSOs]
{{\it XMM-Newton} observations of seven soft X-ray excess QSOs}
\author[K.L. Page \etal]{K.L. Page$^{1}$, N. Schartel$^{2}$, M.J.L. Turner$^{1}$ and P.T. O'Brien$^{1}$\\
$^{1}$ X-Ray and Observational Astronomy Group, Department of Physics \& Astronomy, University of Leicester, LE1 7RH, UK\\
$^{2}$ {\it XMM-Newton} Science Operations Centre, European Space Agency, Villafranca del Castillo, Apartado 50727, E-28080 Madrid, Spain\\
}

\date{Received / Accepted}

\label{firstpage}

\maketitle

\begin{abstract}

{\it XMM-Newton} observations of seven QSOs are presented and the EPIC spectra analysed. Five of the AGN show evidence for Fe K$\alpha$ emission, with three being slightly better fitted by lines of finite width; at the 99~per~cent level they are consistent with being intrinsically narrow, though.
The broad-band spectra can be well modelled by a combination of different temperature blackbodies with a power-law, with temperatures between kT~$\sim$~100--300~eV. On the whole, these temperatures are too high to be direct thermal emission from the accretion disc, so a  Comptonization model was used as a more physical parametrization. The Comptonizing electron population forms the soft excess emission, with an electron temperature of $\sim$~120--680~eV. Power-law, thermal plasma and disc blackbody models were also fitted to the soft X-ray excess. Of the sample, four of the AGN are radio-quiet and three radio-loud. The radio-quiet QSOs may have slightly stronger soft excesses, although the electron temperatures cover the same range for both groups.


\end{abstract}


\section{Introduction}
\label{sec:intro}

At energies below $\sim$~2~keV, the spectra of most AGN show an upturn, away from the extrapolation of the high energy (2--10~keV) power-law. This so-called `soft excess' emission is thought to be common in both Seyfert galaxies and QSOs (Quasi-Stellar Objects). The first such soft excess was identified in Mrk~841 by Arnaud \etal (1985); Turner \& Pounds (1989) then found that $\sim$~50~per~cent of their {\it EXOSAT} sample showed steeper spectral slopes at low energies. Likewise, Walter \& Fink (1993) and Schartel \etal (1996) found that the {\it ROSAT} PSPC spectral index tends to be significantly steeper than that measured above 2.4~keV (typically $\sim$~1.9; e.g., Nandra \& Pounds 1994). More recently, Pounds \& Reeves (2002) have discussed the frequent presence of soft excesses in {\it XMM-Newton} data.

Many papers have been published about the soft excess, covering both observational results -- with {\it ROSAT} (e.g., Fiore \etal 1994; Piro, Matt \& Ricci 1997), {\it Ginga}, {\it EXOSAT} (e.g., Saxton \etal 1993) and {\it Einstein} (e.g., Masnou \etal 1992; Zhou \& Yu 1992) -- and theoretical work (e.g., Czerny \& Elvis 1987; Czerny \& {\. Z}ycki 1994; Xia \& Zhang 2001). It is generally thought that the soft excess may be linked to the hot tail-end of the Big Blue Bump (BBB), or is an extension of the UV band. The BBB/UV excess is likely to be due to thermal emission from the accretion disc surrounding the black hole (e.g., Shields 1978; Malkan \& Sargent 1982). However, this thermal emission is not hot enough to account for the soft X-ray flux as well; hence, Comptonization is often invoked to explain the resultant emission. In this scenario, the direct thermal emission from the accretion disc is observed as the optical and UV spectrum. Some of the disc photons, however, undergo inverse Compton scattering with a population of hot electrons, thus gaining energy and producing a broader spectrum, which appears similar to a power-law over a limited energy range (assuming unsaturated Comptonization).

All seven objects in this paper, listed in Table~\ref{z_wa}, have been previously observed by {\it ROSAT}
(Schartel \etal 1996) and were each noted to have steep photon indices over the 0.1--2.4~keV {\it ROSAT} band. The Galactic absorption in the direction of each low-redshift QSO is small and it has been previously found that there is no significant evidence for additional, intrinsic absorption in any of the objects. These QSOs, therefore, represent a useful sample to investigate the soft excess in both radio-quiet and radio-loud AGN.

\begin{table*}
\begin{center}
\caption{Redshifts and Galactic absorption for the seven objects. Radio data were obtained from NVSS (Condon \etal 1998). Optical data were taken from the NASA Extragalactic Database (NED), while the UV magnitudes were obtained from the Optical Monitor (OM) onboard {\it XMM-Newton} where possible (Mason \etal 2001). The corresponding wavelengths for the different bands are V -- 550~nm; UVW1 -- 291~nm; UVM2 -- 231~nm; UVW2 -- 212~nm} 
\label{z_wa}
$^{a}$ Radio fluxes from the NVSS.
\begin{tabular}{p{1.8truecm}p{0.9truecm}p{1.8truecm}p{1.0truecm}p{1.8truecm}p{1.0truecm}p{1.0truecm}p{1.0truecm}p{1.0truecm}}
\hline
object & RL/RQ & 1.4 GHz &redshift & Gal. abs. & \multicolumn{4}{c}{Optical and UV magnitudes}\\
 & & flux (mJy)$^{a}$ & & (10$^{20}$~cm$^{-2}$) & V-band & UVW1 & UVM2 & UVW2\\
\hline
Q0056$-$363 & RQ & $<$~2.5  &0.162 & 1.93 & 16.7 & --- & --- & 13.9\\ 
PG 0804+761 & RQ & 3.3~$\pm$~0.4 & 0.100 & 2.98 & 15.2 & --- & --- & ---\\
Mrk 1383 & RQ &  2.7~$\pm$~0.5 &0.086 & 2.85 & 17.5 &  --- & 12.4 & 12.4\\
Mrk 876 & RQ & 3.9~$\pm$~0.5 & 0.129 & 2.87 & 15.2 & 13.7 & 13.7 & 13.7\\
B2 1028+31 & RL & 230.5~$\pm$~8 &  0.178 & 1.96 & 16.7 & --- & --- & ---\\
B2 1128+31 & RL & 369.7~$\pm$~13  & 0.289 & 2.02 & 16.6& 15.3 & 14.4 & 14.3\\
B2 1721+34 & RL & 518.3~$\pm$~19.7 & 0.206 & 3.11 & 16.5 & --- & ---  & ---\\
\hline
\end{tabular}
\end{center}
\end{table*}



\section{XMM-Newton Observations}
\label{sec:xmmobs}

The QSOs in this paper were observed by {\it XMM-Newton} between
revolutions 105 and 315 (Table~\ref{obs}). {\sc sas} (Science Analysis Software) v5.4 was used to produce the event
lists for the MOS and PN EPIC (European Photon Imaging Camera) instruments, which were then filtered
using {\sc xmmselect}. For most of the objects, events covering patterns 0--12 (single, double, triple and quadruple events) were selected
for the MOS exposures, while 0--4 (singles and doubles) were used for
the PN, after ensuring that there were no problems with pile-up in either instrument.  PG~0804+761 did appear to show pile-up effects (using the {\sc sas} task {\sc epatplot}), though, so pattern zero spectra (i.e., single events only) were used for this object. Spectra were extracted within a circular region of
$\sim$~40~arcsec centred on each object; background spectra were obtained
from an adjacent area of `blank sky'. Finally, the spectra were binned in such a way that each individual bin contained a sufficient number of counts to apply the $\chi^{2}$ statistic (the minimum used was 25) and the energy range of each bin is larger than one third of the FWHM in order to avoid oversampling effects (i.e. to ensure that the bins were uncorrelated). Version 11.1.0 of {\sc Xspec} was then used to analyse the data.



Table~\ref{z_wa} lists whether the QSOs are radio loud or quiet,
their redshift and Galactic absorption. The 1.4~GHz radio fluxes were taken from the NRAO VLA Sky Survey (NVSS; Condon \etal 1998) catalogue browser. The completeness limit of the survey is $\sim$~2.5~mJy so, since no object was catalogued at the coordinates of Q~0056$-$363, 2.5~mJy is taken as the upper limit to the radio flux. Table~\ref{obs} gives information about the set-up of {\it XMM-Newton} for each observation, while Table~\ref{coords} then gives the Right Ascension and Declination of the objects, together with source count-rates and net counts.

Although the objects will be referred to as QSOs throughout this paper, from their definition in the V{\'e}ron-V{\'e}ron catalogue, their magnitudes are on the border between bright Seyfert galaxies and QSOs (M$_{B}$~$\sim$~$-$23) and they are thus sometimes referred to as Seyfert 1 galaxies in the literature. H$_{0}$ has been taken as 70~km~s$^{-1}$~Mpc$^{-1}$, and q$_{0}$~=~0.5; errors are given at the 1$\sigma$ level (e.g., $\Delta\chi^{2}$~=~2.3 for two interesting parameters).

\begin{table*}
\begin{center}
\caption{Details of the {\it XMM-Newton} observations performed.}
$^{a}$ LW -- large window; SW -- small window; FF -- full frame. 
\label{obs}
\begin{tabular}{p{2.0truecm}p{2.5truecm}p{1.0truecm}p{0.9truecm}p{0.9truecm}p{1.2truecm}p{1.2truecm}p{1.2truecm}p{1.2truecm}p{1.2truecm}p{1.2truecm}p{1.2truecm}}
\hline
object & observation & \multicolumn{3}{c}{exposure time (ks)} & \multicolumn{3}{c}{mode$^{a}$} & \multicolumn{3}{c}{filter} \\
 & date (rev.) & MOS1 & MOS2 & PN & MOS1 & MOS2 & PN & MOS1 & MOS2 & PN\\
\hline

Q0056$-$363 & 2000-07-05 (105) & 5.3/5.3 & 5.3/5.3 & 14.5 & LW & LW & LW & thin/thick & thin/thick & thin\\  
PG 0804+761 & 2000-11-04 (166) & 6.7 & 6.7 & 0.6 & LW & LW & FF & thin & thin & thin\\
Mrk 1383 & 2000-07-28 (116) & 5.0 & 4.9 & 3.5 & LW & LW & LW & thin & thick & thin\\
Mrk 876 & 2001-04-13 (246) & 4.2 & 4.4 & 3.5 & LW & LW & FF & thin & thick & thin\\
& 2001-08-29 (315) & 7.2 & 7.2 & 2.6 & LW & LW & FF & thin & thick & thin\\
B2 1028+31 & 2000-12-06 (182) & 26.2 & 26.2 & 21.5 & LW & LW & FF & thin & thick & thin\\
B2 1128+31 & 2000-11-22 (175) & 23.3 & 23.3 & 18.9 & LW & LW & FF & thin & thick & thin\\
B2 1721+34 & 2001-02-13 (216) & 6.9 & 6.9 & 4.1 & LW & LW & FF & thin & thick & thin\\
 & 2001-02-26 (223) & 6.3 & 6.5 & 3.2 & LW & LW & FF & thin & thick & thin\\  
\hline
\end{tabular}
\end{center}
\end{table*}

\begin{table*}
\begin{center}
\caption{Details of the source positions and extracted spectra. The extraction regions are shown in terms of physical pixels in the X and Y directions and give the centre of the circular region used.} 
$^{a}$ combined counts for the thin filter (MOS1+MOS2) observations; $^{b}$ combined counts for the thick filter (MOS1+MOS2) observations
\label{coords}
\begin{tabular}{p{1.8truecm}p{1.1truecm}p{1.4truecm}p{1.2truecm}p{1.2truecm}p{1.2truecm}p{1.0truecm}p{1.0truecm}p{1.0truecm}p{0.9truecm}p{0.9truecm}p{0.9truecm}}
\hline
Object & RA & Dec & \multicolumn{3}{c}{Extraction region (physical pixels)} & \multicolumn{3}{c}{source count-rates (count s$^{-1}$)} & \multicolumn{3}{c}{net source counts (10$^{3}$)}\\
& & & MOS 1 & MOS 2 & PN & MOS 1 & MOS 2 & PN & MOS 1 & MOS 2 & PN\\
\hline
Q0056$-$363 & 00:58:37.4 & $-$36:06:05.0 & (24240.5, 24800.5) & (24240.5, 24800.5) & (24240.5, 24800.5) & 0.660 $\pm$~0.008$^{a}$ & 0.570 $\pm$~0.007$^{b}$ & 3.178 $\pm$~0.015 & 7.1 & 6.1 & 46.5\\
PG 0804+761 & 08:10:58.5 & 76:02:43.0 & (24600.5, 24240.5) & (24560.5, 24280.5) & (24600.5, 24360.5) & 1.877 $\pm$~0.017 & 1.962 $\pm$~0.017 & 7.888 $\pm$~0.119 & 12.8 & 13.3 & 14.5\\
Mrk 1383 & 14:29:06.6 & 01:17:06.0 & (26520.5, 27840.0) & (26520.5, 27840.0) & (26520.5, 27840.0) & 1.866 $\pm$~0.019 & 1.633 $\pm$~0.018 & 8.718 $\pm$~0.063 & 9.5 & 8.1 & 19.7\\
Mrk 876 & 16:13:57.2 & 65:43:09.0 & (23880.5, 25680.5) & (23920.5, 25720.5) & (23960.5, 25720.5) & 0.821 $\pm$~0.009 & 0.706 $\pm$~0.008 & 3.328 $\pm$~0.026 & 9.8 & 8.3 & 23.2\\
B2 1028+31 & 10:30:59.1 & 31:02:56.0  & (25280.5, 23960.5) & (25320.5, 24000.5) & (25280.5, 23960.5) & 0.404 $\pm$~0.004 & 0.351 $\pm$~0.003 & 1.708 $\pm$~0.009 & 10.7 & 9.3 & 37.3\\
B2 1128+31 & 11:31:09.4 & 31:14:07.0 & (25720.5, 23840.5) & (25720.5, 23840.5) & (25560.5, 24120.5) & 0.410 $\pm$~0.004 & 0.366 $\pm$~0.004 & 1.871 $\pm$~0.011 & 9.7 & 8.6 & 37.6\\ 
B2 1721+34 &  17:23:20.8 & 34:17:59.0 & (25680.5, 23880.5) & (25680.5, 23840.5) & (25761.0, 23921.0) & 1.349 $\pm$~0.011 & 1.195 $\pm$~0.010 & 5.138 $\pm$~0.028 & 18.4 & 16.1 & 38.5\\

\hline
\end{tabular}
\end{center}
\end{table*}

\section{Spectral Analysis}
\label{sec:specanal}

\subsection{2--12~keV}

\begin{figure*}
\begin{center}
\includegraphics[clip,height=11cm,angle=-90]{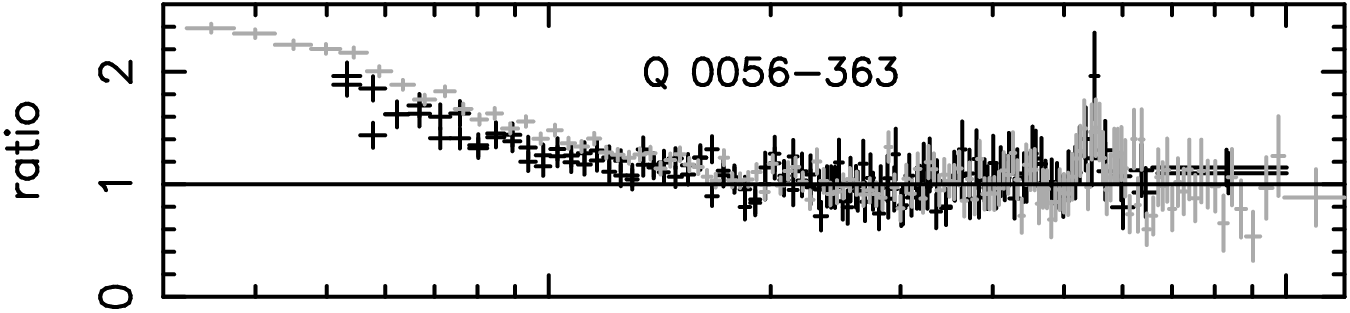}
\includegraphics[clip,height=11cm,angle=-90]{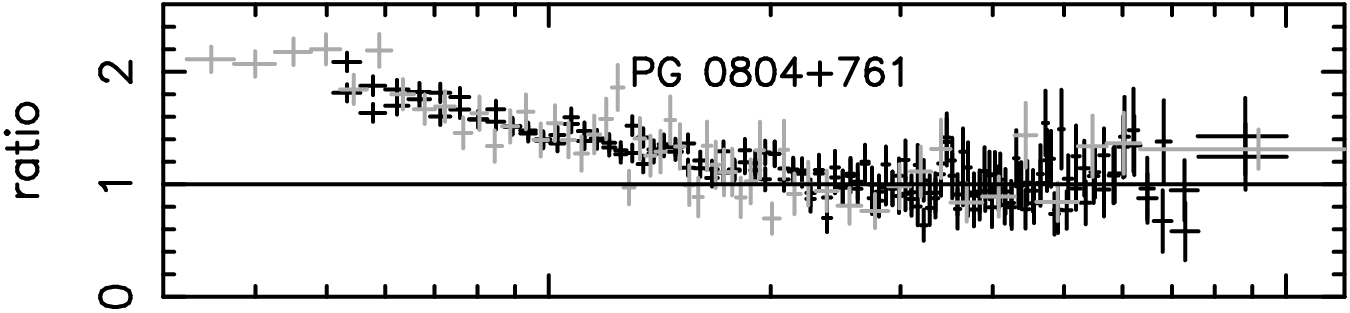}
\includegraphics[clip,height=11cm,angle=-90]{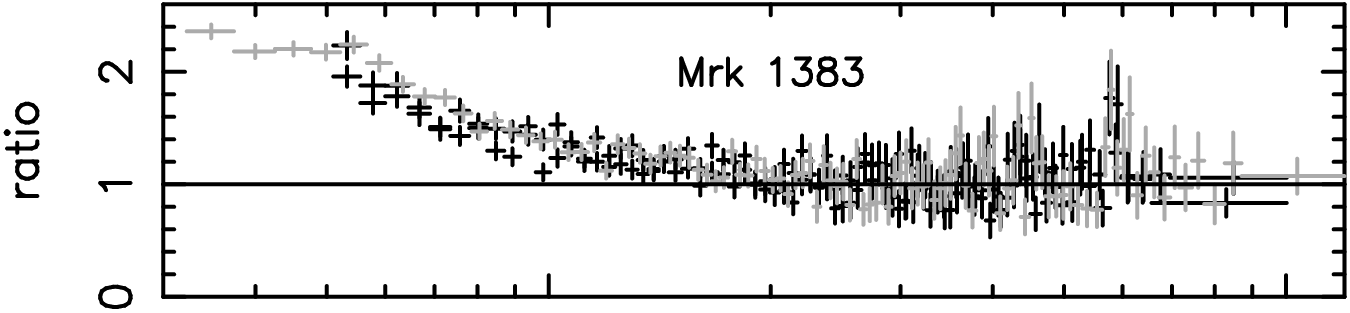}
\includegraphics[clip,height=11cm,angle=-90]{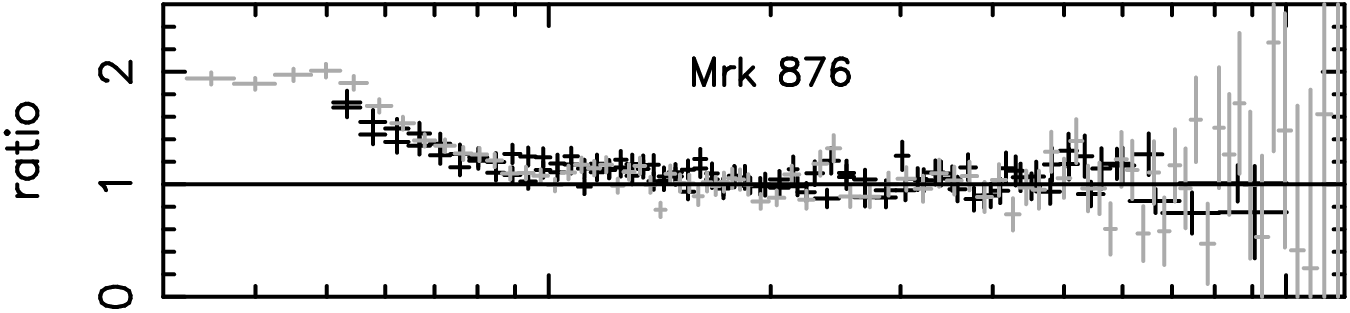}
\includegraphics[clip,height=11cm,angle=-90]{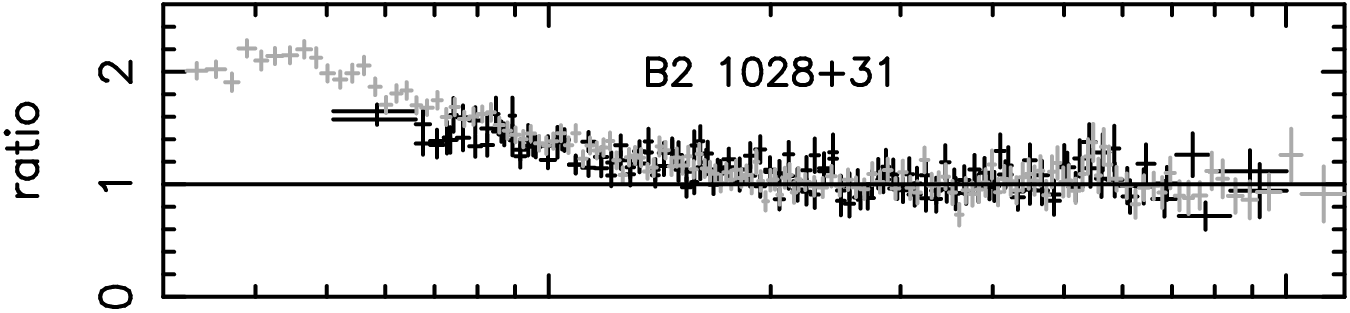}
\includegraphics[clip,height=11cm,angle=-90]{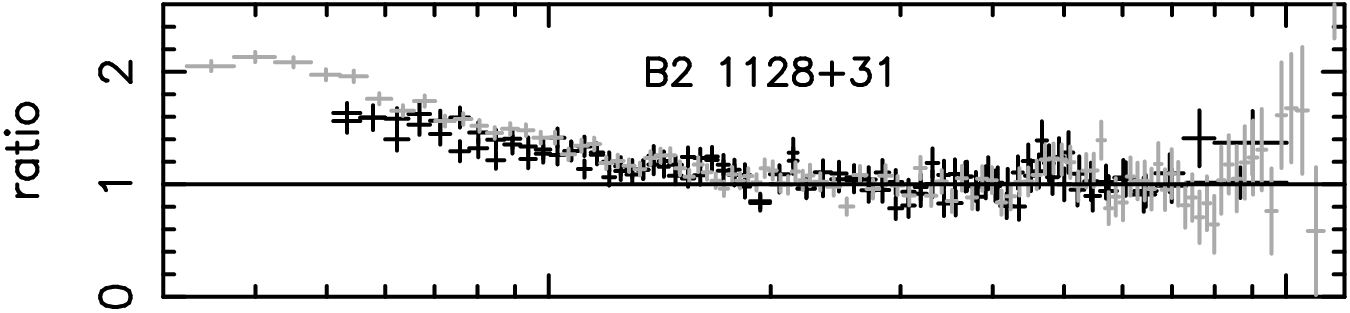}
\includegraphics[clip,height=11cm,angle=-90]{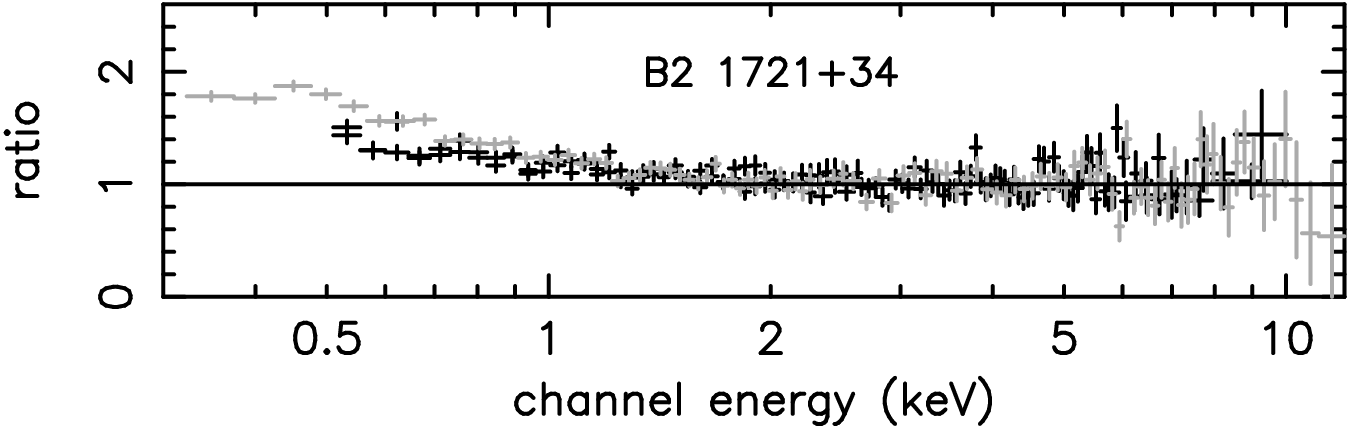}
\caption{Ratio plots of the power-law fits above 2~keV, extrapolated down to lower energies, to show the soft excesses. The MOS data-sets are shown in black, the PN, in grey.}
\label{lines}
\end{center}
\end{figure*}

\begin{figure*}
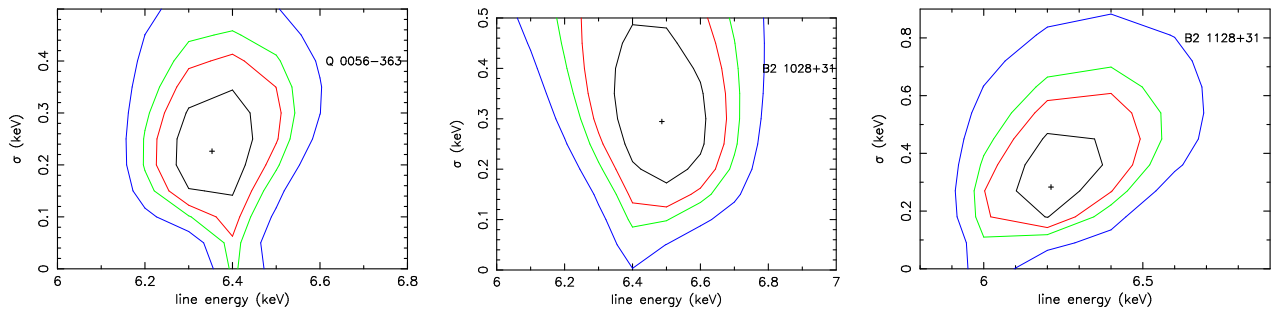

\begin{center}
\includegraphics[clip,width=4.0cm,angle=-90]{Q0056_varbin_cont.cps}\hspace*{0.3cm}
\includegraphics[clip,width=4.0cm,angle=-90]{1028_varbin_cont.cps}\hspace*{0.3cm}
\includegraphics[clip,width=4.0cm,angle=-90]{1128_varbin_cont.cps}
\caption{68, 90, 95 and 99~per~cent confidence contours for the energy and width of the iron emission lines.}
\label{cont}
\end{center}
\end{figure*}

As is conventional, each background-subtracted spectrum was first simply fitted with a
single power-law over the entire band pass of {\it
XMM-Newton}, attenuated by the Galactic absorption column relevant for each object (Table~\ref{z_wa}). MOS1, MOS2 and PN gave consistent results, so the spectra were fitted simultaneously; the values given in the tables are those for the joint fits. It was found, however, that a power-law over the broad-band produced bad fits for all of
the spectra, generally due to upward curvature at energies lower than
$\sim$~1--2~keV. Therefore, the
fits were repeated above 2~keV (observer's frame) for each object, up to 10~keV for the MOS cameras and 12~keV for PN. Many showed positive residuals between
$\sim$~6--7~keV (Figure~\ref{lines}), indicative of iron emission lines, so Gaussian components
were added. Initially, the width of the line was constrained to be
unresolved by the EPIC instruments (i.e., $\sigma$~=~ 0.01~keV); the width was then allowed to vary for the following fit. Table~\ref{hard} shows the results to the high energy fits for the QSOs, along with F-test null probabilities for the inclusion of the narrow and broad Gaussian lines. The line widths were found to be unconstrained for Mrk~876 and B2~1721+34, so the 1$\sigma$ upper limit has been given in the table. If the F-test null probability is $>$ 1~$\times$~10$^{-2}$, then the line is less than 99~per~cent significant; in this case, the 1$\sigma$ upper limit on the equivalent width (EW) has been quoted. For those QSOs (Q~0056$-$363, B2~1028+31 and B2~1128+31) which appear to be better fitted by lines of finite width, contour plots showing the relationship between the line energy and width are given in Figure~\ref{cont}; the contours plotted are for 68, 90, 95 and 99~per~cent confidence levels. It can be seen that, at the 99~per~cent level, the width of the lines is consistent with being intrinsically narrow.

\begin{table*}
\begin{center}
\caption{Fits over 2--10 (MOS) and 2--12~keV (PN) in the observer's frame, where $\Gamma$ is the power-law slope. NGA~=~narrow Gaussian line, with width fixed at 0.01~keV; FGA~=~free Gaussian line, with width allowed to vary.  The penultimate column gives the F-test null probability for that fit compared to the previous one for that object. If $\chi^{2}$ does not change, then the F-value is undefined.} 
$^{a}$ energy given in the rest frame of the QSO; $^{b}$ intrinsic width of line; $^{f}$ frozen
\label{hard}

\begin{tabular}{p{2.0truecm}p{2.0truecm}p{1.5truecm}p{1.6truecm}p{1.6truecm}p{1.5truecm}p{1.0truecm}p{1.8truecm}p{2.0truecm}}
\hline
object & model &  $\Gamma$ & line energy$^{a}$  & $\sigma$$^{b}$
 & EW & $\chi^{2}$/dof & F-test null & 2-10 keV lum.\\
 & & & (keV) &(keV) &(eV) &  & probability & (10$^{44}$ erg s$^{-1}$)\\
\hline
Q 0056$-$363 &  PL & 1.98~$\pm$~0.03 & & & &  162/171\\   
 & PL + NGA & 2.00~$\pm$~0.03 & 6.43~$\pm$~0.03 & 0.01$^{f}$  & 156~$\pm$~35 & 143/169 & 2.64~$\times$~10$^{-5}$ \\
 & PL + FGA & 2.03~$\pm$~0.04 & 6.35~$\pm$~0.06 & 0.23~$\pm$~0.07 & 276~$\pm$~68 & 137/168 & 7.37~$\times$~10$^{-3}$  & 1.79\\
&\\
PG 0804+761   & PL & 1.93~$\pm$~0.05 & & & &  115/122\\
 & PL + NGA & 1.96~$\pm$~0.05 & 6.73~$\pm$~0.04 & 0.01$^{f}$ & $<$263 & 107/120 & 1.32~$\times$~10$^{-2}$ \\
 & PL + FGA & 2.00~$\pm$~0.06 & 6.62~$\pm$~0.14 & 0.33~$\pm$~0.16 & $<$512 & 106/119 & 0.291 & 2.54 \\
&\\
Mrk 1383  & PL  & 1.96~$\pm$~0.04 & & & &  166/160\\
 & PL + NGA &  1.98~$\pm$~0.04 & 6.31~$\pm$~0.03 & 0.01$^{f}$ & 140~$\pm$~41 & 153/158 & 1.59~$\times$~10$^{-3}$ \\
 & PL + FGA & 1.99~$\pm$~0.04 & 6.39~$\pm$~0.06 & 0.12~$\pm$~0.07 & 179~$\pm$~61 & 153/157 & -- & 1.48\\
&\\
Mrk 876 & PL & 1.83~$\pm$~0.04 & & & & 213/242\\
 & PL + NGA & 1.84~$\pm$~0.04 & 6.51~$\pm$~0.13 & 0.01$^{f}$ & $<$73 & 212/240 & 0.569 \\
 & PL + FGA & 1.84~$\pm$~0.04 & 6.49~$\pm$~0.17 & $<$0.18 & $<$85 & 212/239 & -- & 1.82\\
&\\
B2 1028+31  & PL  & 1.67~$\pm$~0.03 & & & &  265/280\\
 & PL + NGA & 1.68~$\pm$~0.03 & 6.43~$\pm$~0.04 & 0.01$^{f}$ & 77~$\pm$~18 & 254/278 & 2.76~$\times$~10$^{-3}$ \\ 
 & PL + FGA & 1.71~$\pm$~0.03 & 6.49~$\pm$~0.09 & 0.30~$\pm$~0.10 & 200~$\pm$~57 & 245/277 & 1.59~$\times$~10$^{-3}$ & 1.90\\
&\\
B2 1128+31 & PL & 1.70~$\pm$~0.03 & & & &  244/235\\
 & PL + NGA & 1.72~$\pm$~0.03 & 6.31~$\pm$~0.05 & 0.01$^{f}$ & $<$87 & 236/233 & 2.06~$\times$~10$^{-2}$ \\
 & PL + FGA & 1.74~$\pm$~0.03 & 6.22~$\pm$~0.10 & 0.28~$\pm$~0.11 & 183~$\pm$~59 & 225/232 & 8.87~$\times$~10$^{-4}$ & 5.16\\
&\\
B2 1721+34 & PL  & 1.76~$\pm$~0.02 & & & &  242/263\\
 & PL + NGA & 1.78~$\pm$~0.03 & 6.43~$\pm$~0.05 & 0.01$^{f}$ & 54~$\pm$~20 & 233/261 & 7.11~$\times$~10$^{-3}$ \\
 & PL + FGA & 1.77~$\pm$~0.03 & 6.45~$\pm$~0.05 & $<$0.05 & 56~$\pm$~19 & 233/260 & -- & 8.83\\

\hline
\end{tabular}
\end{center}
\end{table*}

It is well documented (e.g., Wilkes \& Elvis 1987;  Williams \etal 1992; Lawson \etal 1992; Brinkmann, Yuan \& Siebert 1997; Reeves \etal 1997) that Radio-Loud Quasars (RLQs) have flatter spectral indices than their radio-quiet counterparts (RQQs). Although the current sample consists of only four RQQs and three RLQs, the slopes confirm this trend, with the mean values being $\Gamma_{\rm RQQ}$~=~1.97~$\pm$~0.02 and $\Gamma_{\rm RLQ}$~=~1.74~$\pm$~0.02.


\subsection{The soft excess}
\label{sec:softexcess}

To examine the broad-band continuum of these QSOs, the power-law was extrapolated down to 0.5/0.3~keV for MOS/PN respectively. In each case the data points were found to lie well above the power-law, indicating the existence of soft excesses (see Figure~\ref{lines}). In the following sections, various methods of modelling the soft excess are tested.


\subsubsection{Power-law}

Table~\ref{pofits} gives the results of fitting a double power-law model to the datasets. Although the simple power-law fit to the data above 2~keV is generally accepted to give a good indication of the underlying power-law over the broader X-ray energies, these double power-law fits do not always agree with those results. In particular, fitting PG~0804+761 with two power-laws gives a very flat (though unconstrained) value for the power-law slope at higher energies ($\Gamma$~$\sim$~0.23). Fixing the slope to a more reasonable value of $\Gamma$~=~1.9 gives a worse fit, with $\chi^{2}$/dof~=~246/218, an increase of 13, for one additional degree of freedom.

\begin{table}
\begin{center}
\caption{Double power-law fits to the broad-band data. $\Gamma_{se}$ gives the slope of the power-law modelling the soft excess, while $\Gamma_{he}$ gives the slope at higher energies.} 
\label{pofits}
\begin{tabular}{p{2.0truecm}p{2.0truecm}p{2.0truecm}p{1.0truecm}}
\hline
object & $\Gamma_{se}$ & $\Gamma_{he}$ & $\chi^{2}$/dof\\
\hline
Q 0056$-$363 &  2.85~$\pm$~0.07 & 1.60~$\pm$~0.13 & 296/266\\
PG 0804+761   & 2.41~$\pm$~0.03 & 0.23~$\pm$~0.46 & 231/217\\
Mrk 1383  & 2.91~$\pm$~0.11 & 1.66~$\pm$~0.13 & 297/256\\
Mrk 876 & 3.48~$\pm$~0.17 & 1.71~$\pm$~0.05 & 417/341\\
B2 1028+31  & 2.41~$\pm$~0.08 & 1.29~$\pm$~0.15 & 504/460\\
B2 1128+31 & 2.42~$\pm$~0.07 & 1.19~$\pm$~0.16 & 364/330\\
B2 1721+34 & 2.78~$\pm$~0.14 & 1.59~$\pm$~0.08 & 441/360\\

\hline
\end{tabular}
\end{center}
\end{table}

\subsubsection{Blackbody components}
 
\begin{figure*}
\begin{center}
\includegraphics[clip,width=3.8cm,angle=-90]{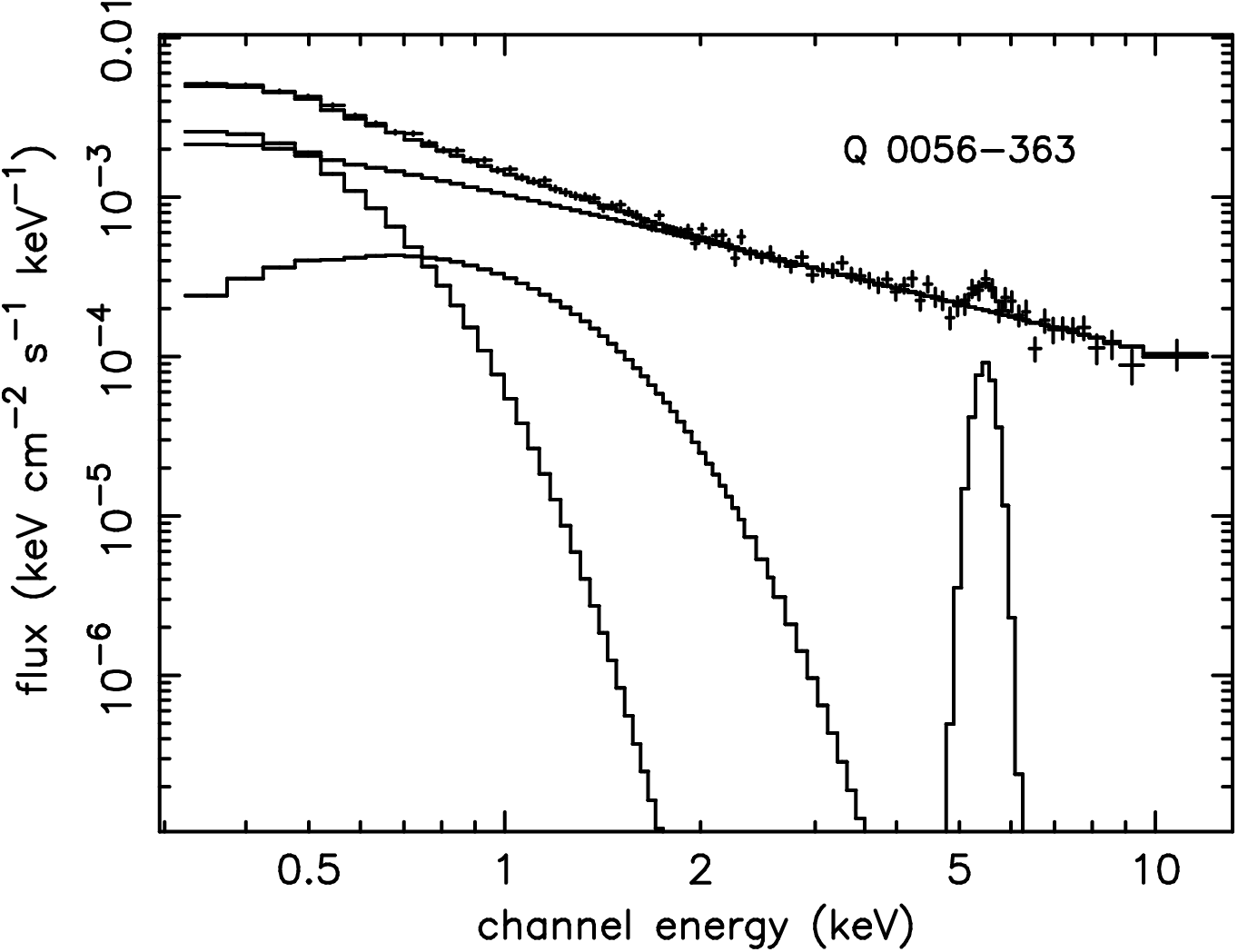}\hspace*{2.0cm}
\includegraphics[clip,width=3.8cm,angle=-90]{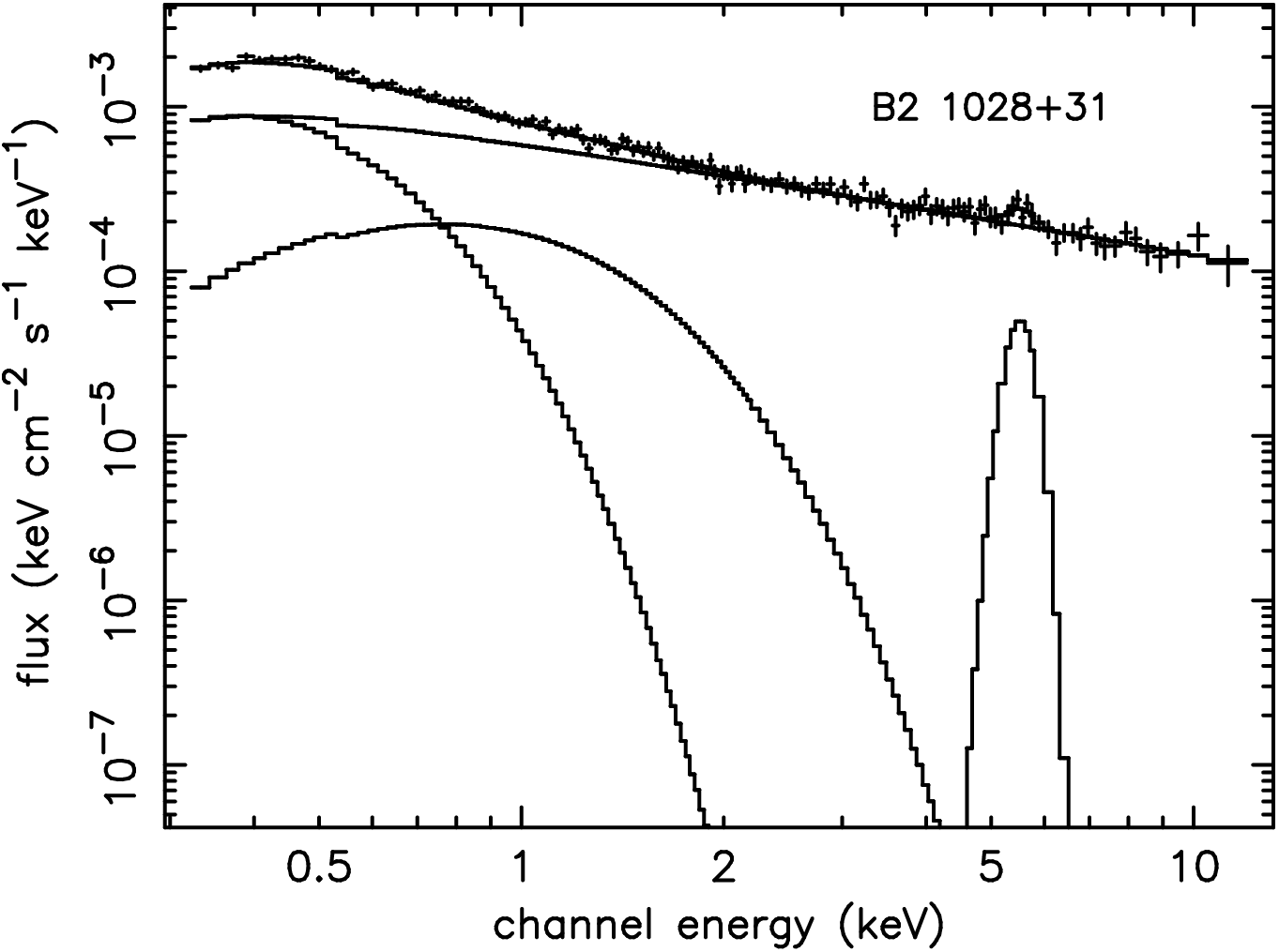}\vspace*{0.01cm}
\includegraphics[clip,width=1.79cm,angle=-90]{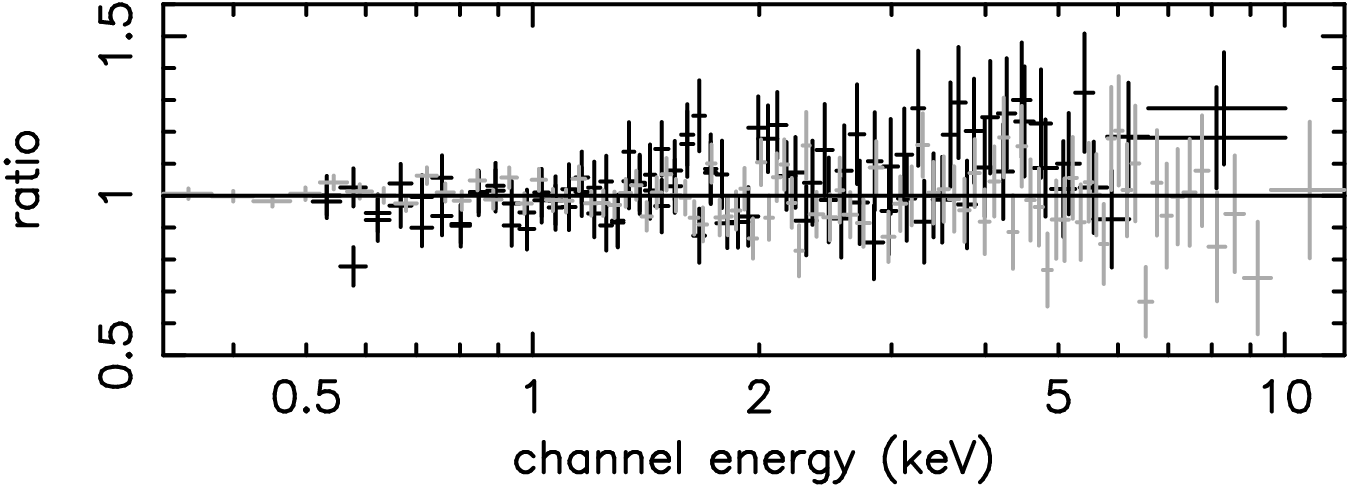}\hspace*{2.0cm}
\includegraphics[clip,width=1.85cm,angle=-90]{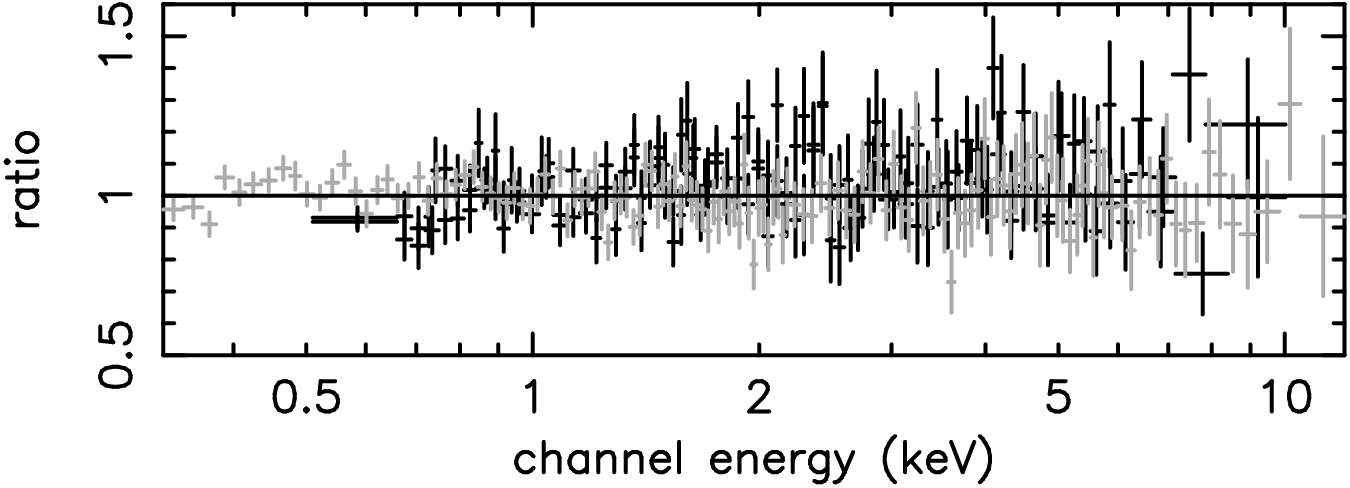}\vspace*{0.1cm}

\includegraphics[clip,width=3.8cm,angle=-90]{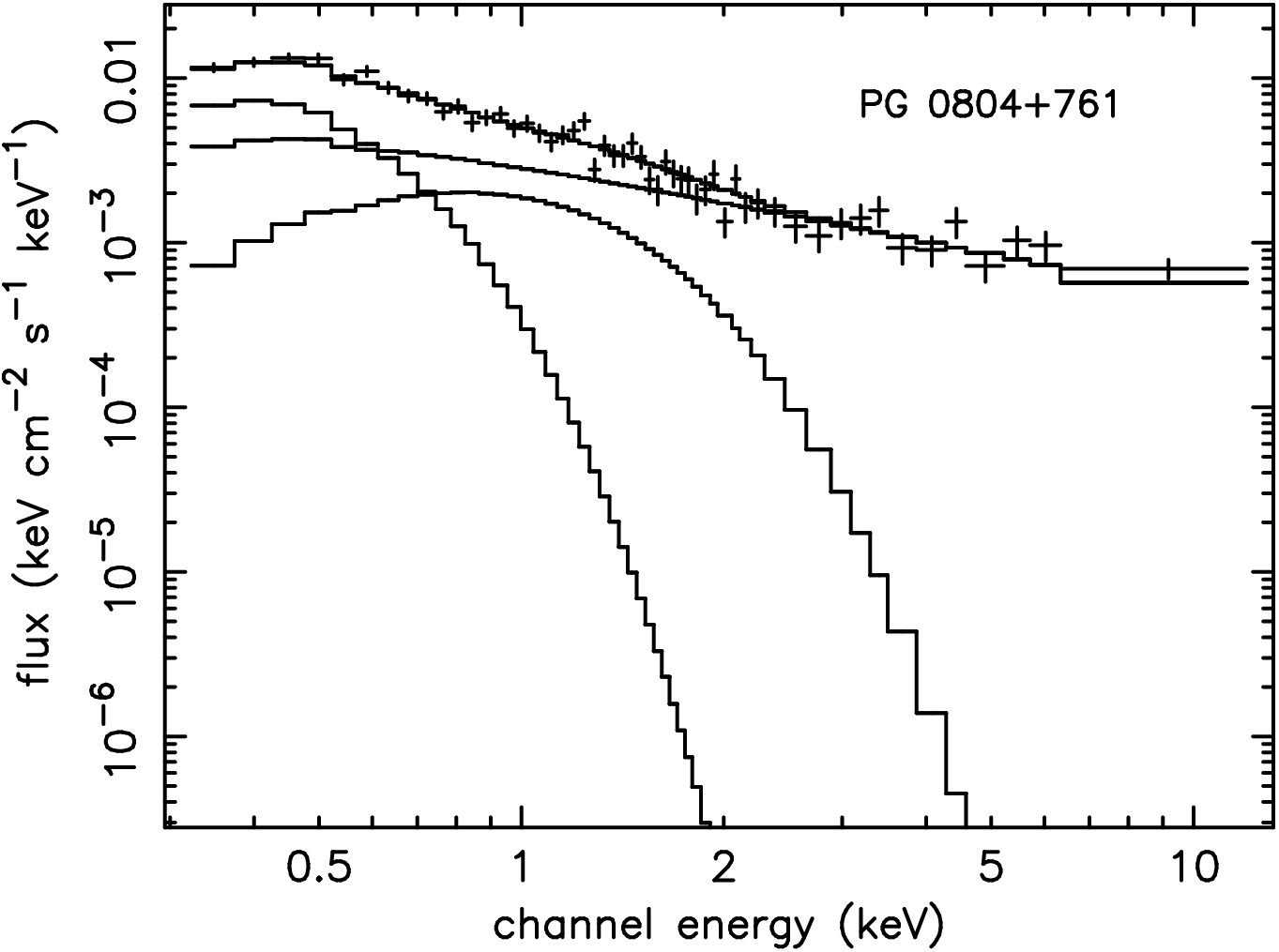}\hspace*{2.0cm}
\includegraphics[clip,width=3.8cm,angle=-90]{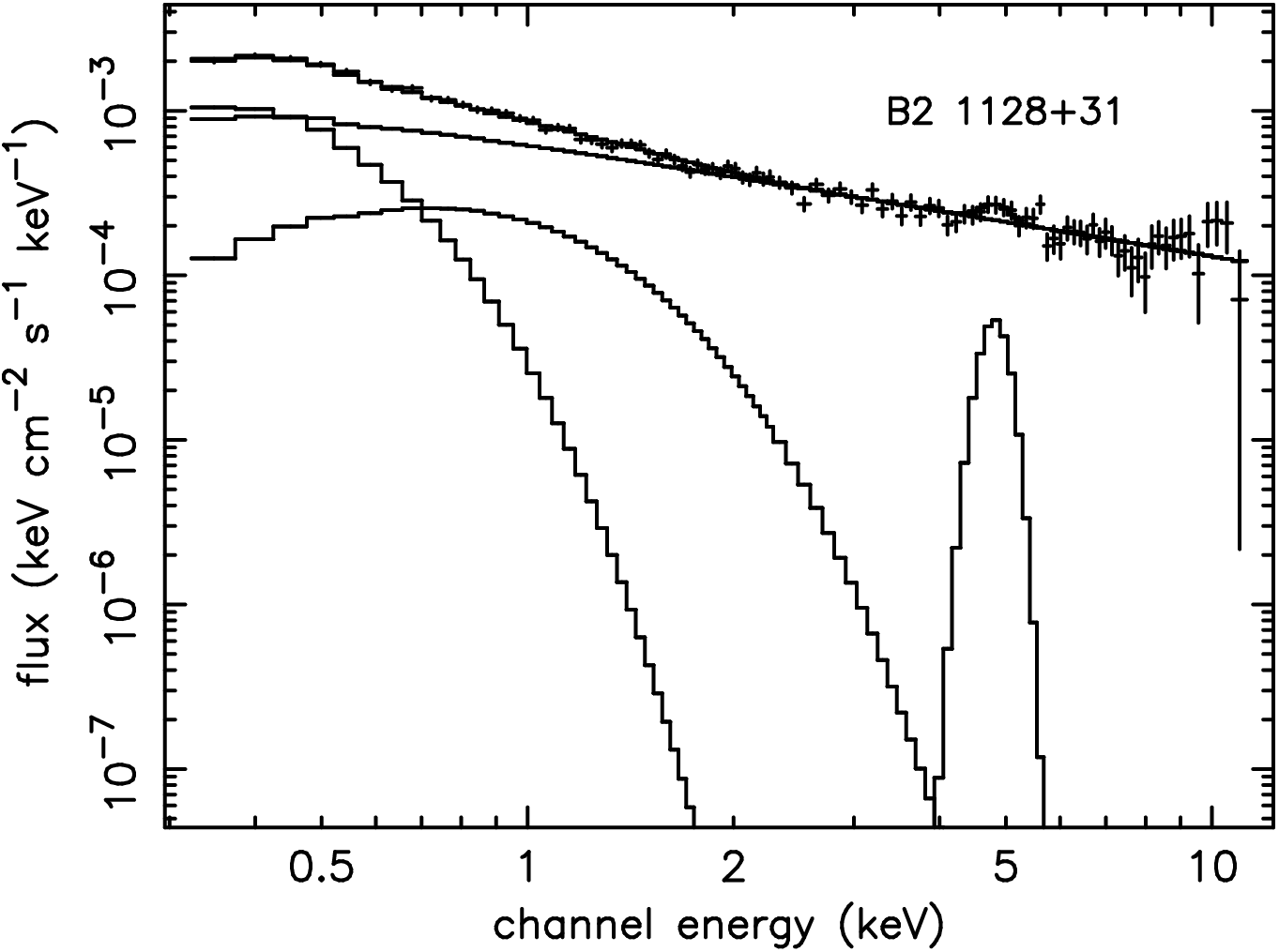}\vspace*{0.01cm}
\includegraphics[clip,width=1.79cm,angle=-90]{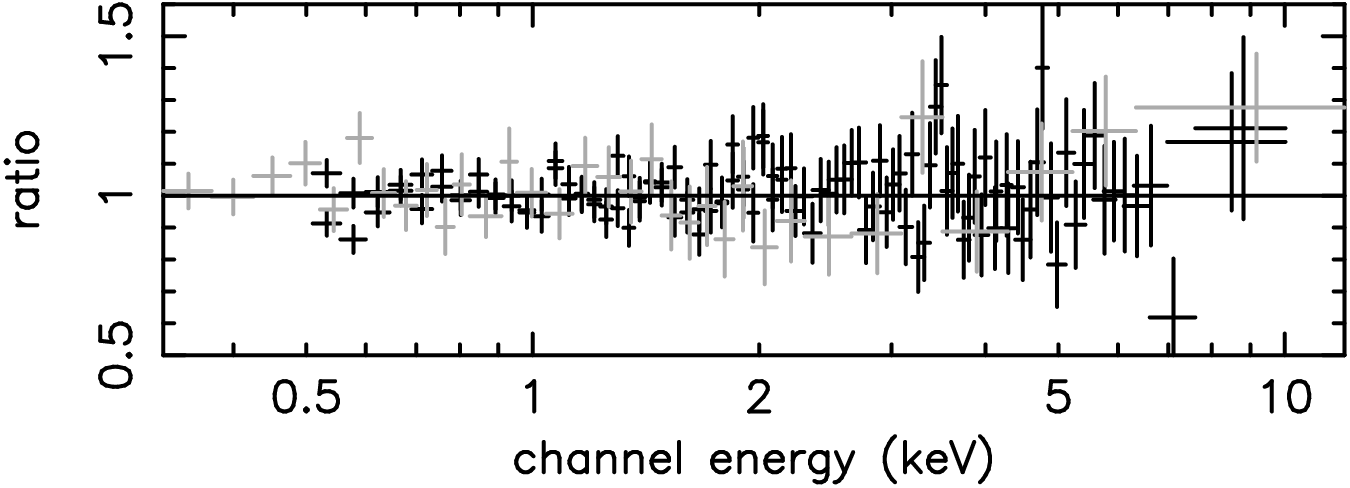}\hspace*{2.0cm}
\includegraphics[clip,width=1.85cm,angle=-90]{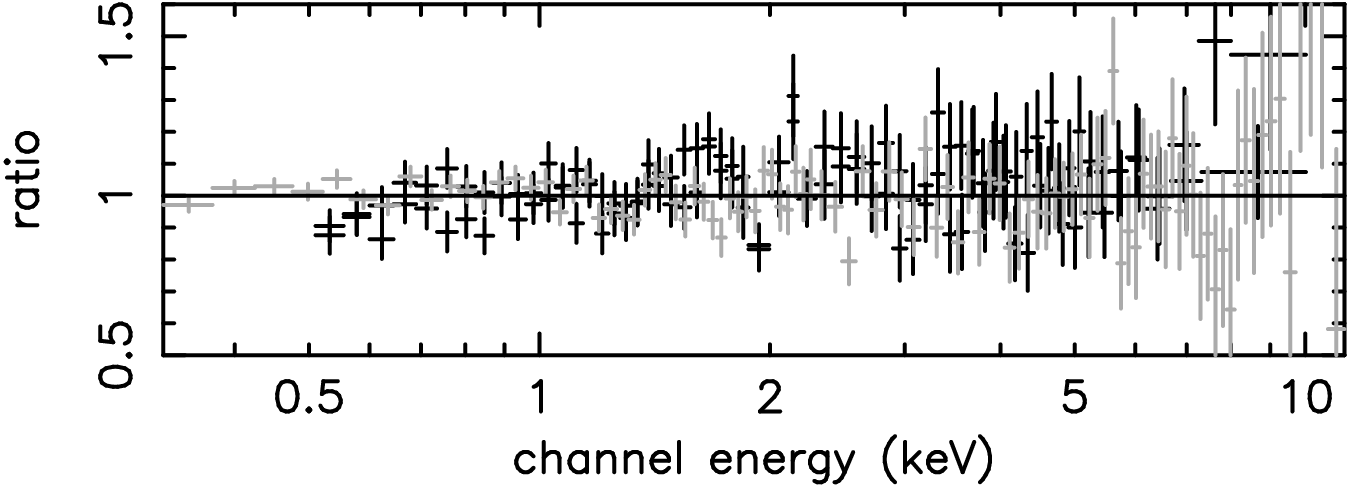}\vspace*{0.1cm}

\includegraphics[clip,width=3.8cm,angle=-90]{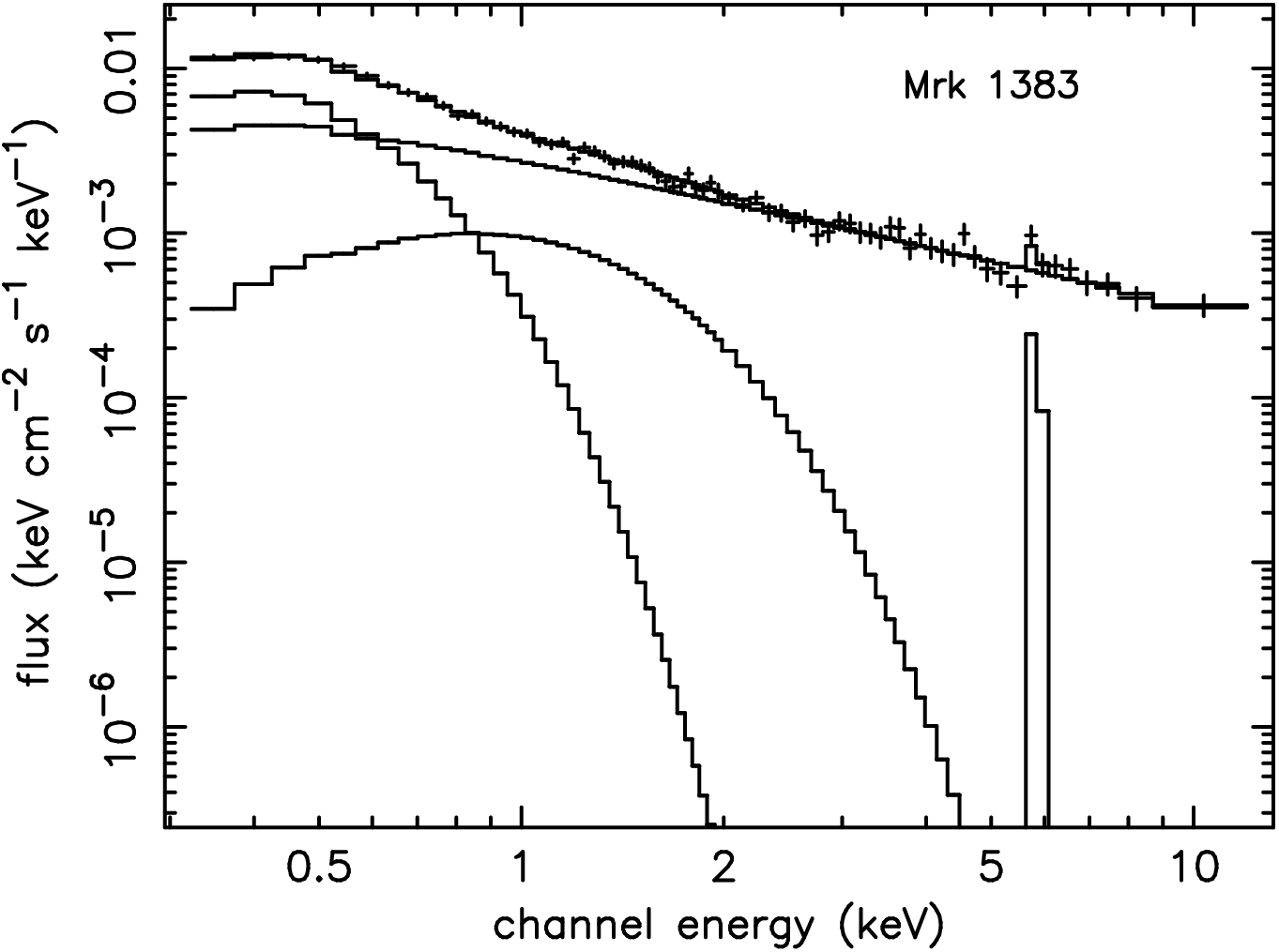}\hspace*{2.0cm}
\includegraphics[clip,width=3.8cm,angle=-90]{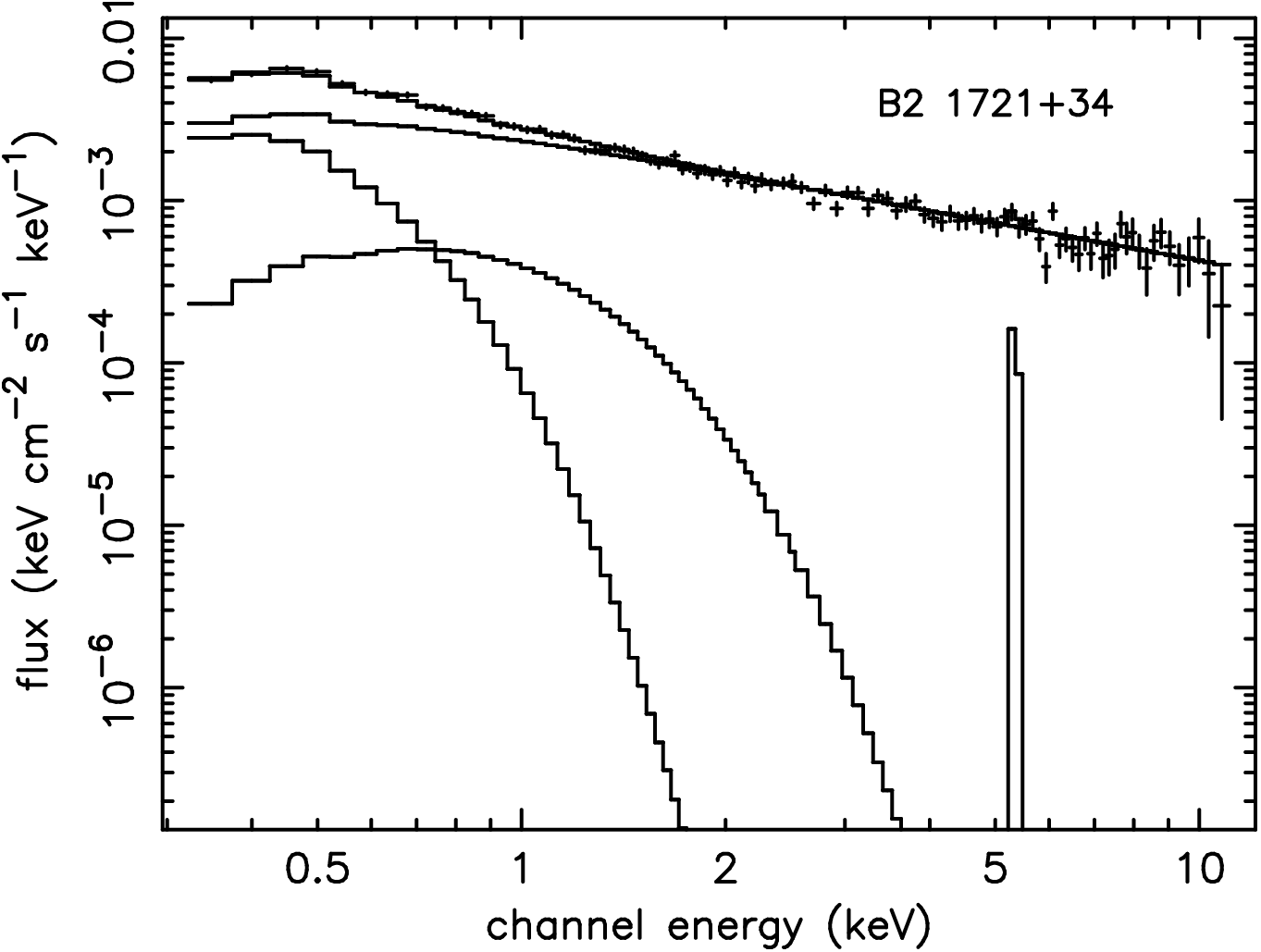}\vspace*{0.01cm}
\includegraphics[clip,width=1.79cm,angle=-90]{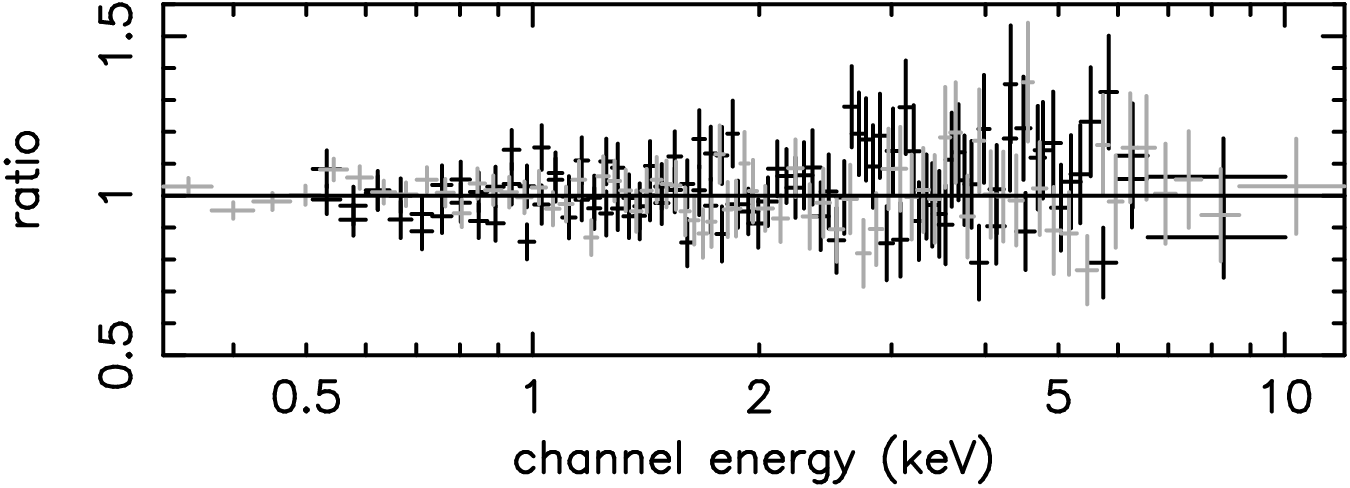}\hspace*{2.0cm}
\includegraphics[clip,width=1.85cm,angle=-90]{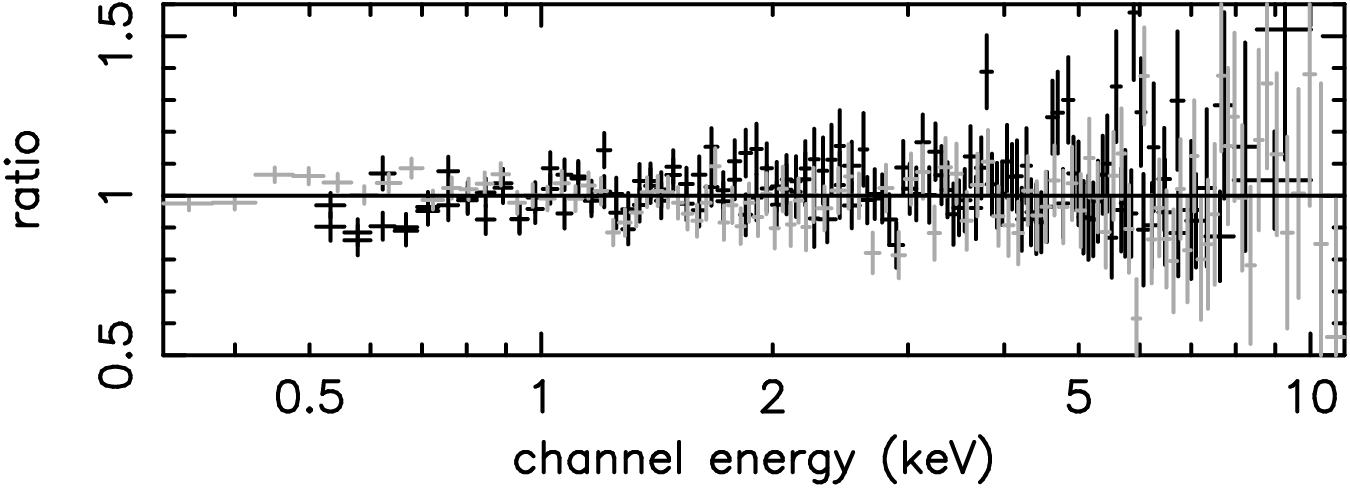}\vspace*{0.1cm}

\includegraphics[clip,width=4.0cm,angle=-90]{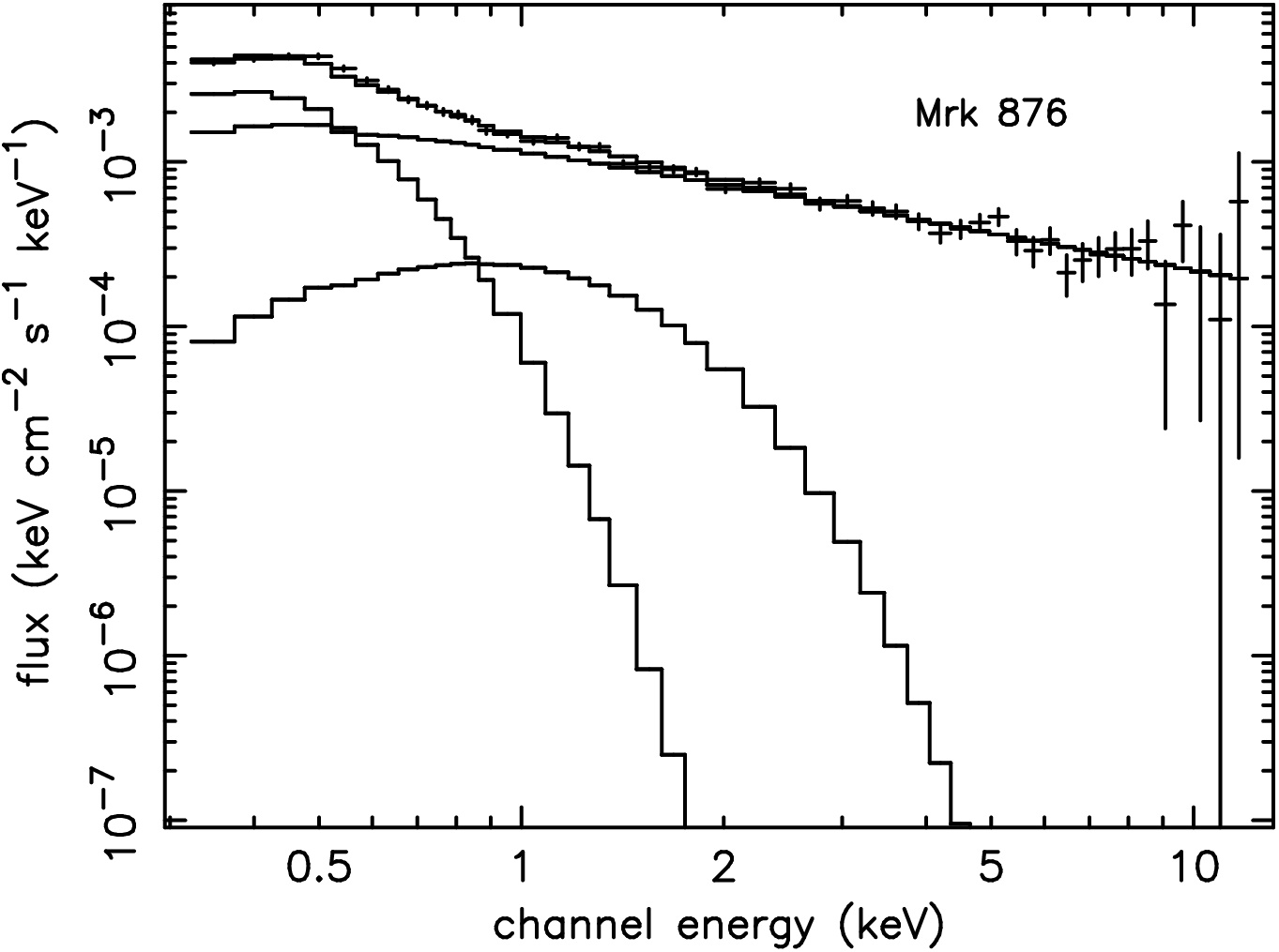}\vspace*{0.01cm}
\hspace{7.0cm}
\includegraphics[clip,width=1.9cm,angle=-90]{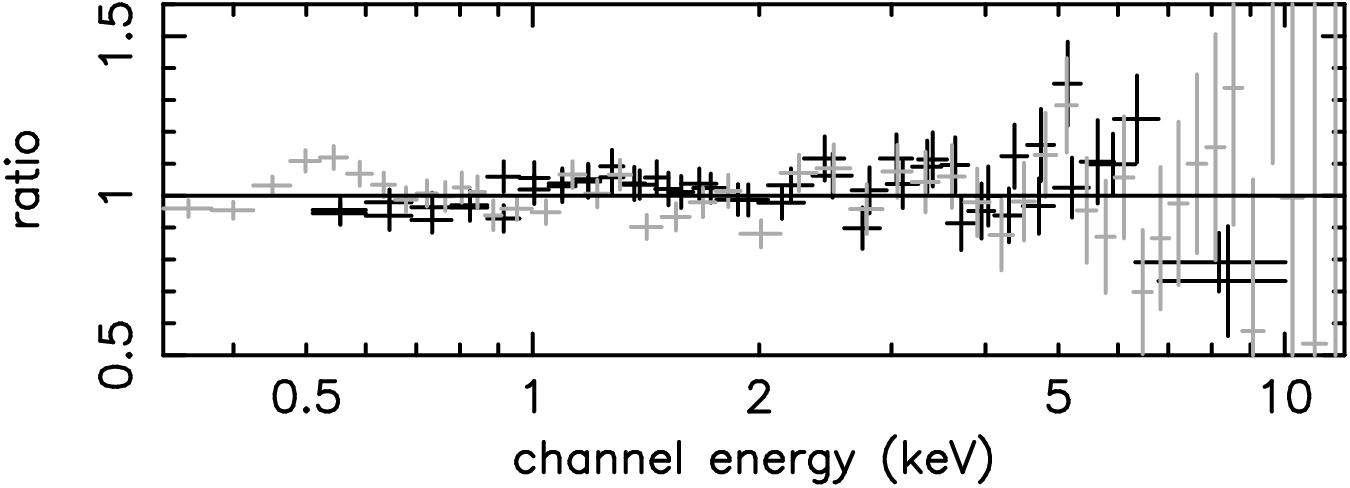}
\caption{Unfolded plots showing the blackbody fits to the soft excess. PN data only are shown for simplicity in the unfolded plots. Both MOS (black) and PN (grey) residuals are shown below each plot.}
\label{bb_euf}
\end{center}
\end{figure*}

Next, the soft excess for each object was parametrized with
multiple blackbody (BB) components; the resulting fits are listed in
Table~\ref{bb} and the unfolded plots are shown in Figure~\ref{bb_euf}. Column 6 of the table gives the ratio of the combined BB luminosity to that of the power-law, over the 0.5--10~keV bandpass. This can be thought of as the `strength' of the soft excess, with larger values indicating that the soft excess is relatively luminous compared to the power-law component. This will be investigated further in Section~\ref{sec:disc}. Using the F-test, with the exception of PG~0804+761 and B2~1128+31, the blackbody fits are preferred at $>~$99~per~cent compared to the power-law parametrization of the soft excess. The two anomalous results are those spectra which are modelled with extremely flat high-energy photon indices; these values for $\Gamma$ do not agree with the simple power-law fits above 2~keV.


\begin{table*}
\begin{center}
\caption{Blackbody fits to the soft excess. The fits
also include iron lines, with energies and widths fixed at the values
previously found (Table~\ref{hard}). The luminosities of the BB components are calculated over the observed 0.5--10~keV band. } 
\label{bb}
\begin{tabular}{p{1.8truecm}p{1.5truecm}p{2.0truecm}p{2.5truecm}p{2.8truecm}p{1.8truecm}p{1.0truecm}}
\hline
object & $\Gamma$ & rest frame kT  & luminosity  of BB & total 0.5--10 keV & BB/$\Gamma$ & $\chi^{2}$/dof \\
 & &(keV) & (10$^{44}$ erg s$^{-1}$) & lum. (10$^{44}$ erg s$^{-1}$) & lum. ratio\\

\hline
Q 0056$-$363 & 2.01~$\pm$~0.04 & 0.105~$\pm$~0.004 & 0.33~$\pm$~0.03 & 3.98 & 0.21 & 272/264 \\
 & & 0.251~$\pm$~0.021 & 0.36~$\pm$~0.06 \\
&\\
PG 0804+761 & 1.85~$\pm$~0.08 & 0.106~$\pm$~0.006 & 0.48~$\pm$~0.07 & 5.47 & 0.31 & 229/215 \\ 
& & 0.289~$\pm$~0.017 & 0.80~$\pm$~0.17\\ 
&\\
Mrk~1383 & 1.91~$\pm$~0.06 & 0.107~$\pm$~0.004 & 0.37~$\pm$~0.03 & 3.22 & 0.27 & 269/254 \\
 & & 0.300~$\pm$~0.022 & 0.32~$\pm$~0.10\\ 
&\\
Mrk~876 & 1.75~$\pm$~0.06 & 0.103~$\pm$~0.004 & 0.28~$\pm$~0.02 & 3.43 &  0.16 & 366/339 \\
& & 0.319~$\pm$~0.035 & 0.19~$\pm$~0.01\\
&\\
B2~1028+31 & 1.69~$\pm$~0.03 & 0.118~$\pm$~0.006 & 0.18~$\pm$~0.02 & 3.44 & 0.14 & 480/458 \\
 & & 0.296~$\pm$~0.026 & 0.24~$\pm$~0.04 \\
&\\
B2~1128+31 & 1.69~$\pm$~0.04 & 0.119~$\pm$~0.006 & 0.47~$\pm$~0.04 & 9.63 & 0.15 & 361/328 \\
 & & 0.304~$\pm$~0.023 & 0.79~$\pm$~0.11 \\
&\\
B2~1721+34 & 1.77~$\pm$~0.03 & 0.109~$\pm$~0.007 & 0.64~$\pm$~0.06  & 16.07 & 0.09 & 421/358 \\
 & & 0.283~$\pm$~0.070 & 0.73~$\pm$~0.15\\
\hline
\end{tabular}
\end{center}
\end{table*}

\subsubsection{Thermal Plasma}

An alternative suggestion for the soft excess is an optically thin thermal plasma model. Since the soft excesses of these QSOs appear to be smooth, it might be expected that the model required would be closer to optically thin emission (Bremsstrahlung) from hydrogen only, which produces a featureless continuum. Table~\ref{brems} gives the results obtained from fitting the spectra with a power-law together with the {\sc xspec} {\it mekal} model (Mewe, Gronenschild \& van den Oord 1986; Mewe, Lemen \& van den Oord 1986; Liedahl, Osterheld \& Goldstein 1995), which produces an emission spectrum from hot, diffuse gas. With the exception of Mrk~876, the spectral fits with the thermal plasma model are worse than those with two blackbody components (F-test probabilities imply that the blackbody fit is better at $>$~99~per~cent). Mrk~876, however, is slightly better fitted with the thermal model.

\subsubsection{Disc blackbody}

Another possibility for the soft emission is a disc blackbody (see, e.g., Mitsuda \etal 1984; Makishima \etal 1986), which models the emission from the accretion disc as a series of blackbodies at different temperatures, emitted from different radii. As Table~\ref{brems} shows,  this model was unable to account for the full breadth of the soft excess, with F-test null probabilities of $<$~5~$\times$~10$^{-5}$ indicating that the multiple blackbodies were preferred at $>$99.99~per~cent.

\begin{table*}
\begin{center}
\caption{Thermal plasma and disc blackbody fits to the QSO spectra. When using the {\it mekal} thermal model, the abundance is given as $\sim$~0 if the value is $<$1~$\times$~10$^{-5}$. The plasma density was fixed at 10$^{12}$ cm$^{-3}$, although the spectra were insensitive to the value.} 
\label{brems}
\begin{tabular}{p{2.0truecm}p{2.0truecm}p{2.0truecm}p{2.0truecm}p{2.0truecm}p{1.0truecm}}
\hline
object & model & $\Gamma$ & kT (keV) & abundance & $\chi^{2}$/$\nu$\\
\hline
Q 0056$-$363 & PL+THERM & 2.08~$\pm$~0.03 & 0.321~$\pm$~0.018 & $\sim$~0 & 283/265\\
 & PL+DISCBB & 2.16~$\pm$~0.02 & 0.140~$\pm$~0.003 & & 294/266\\ 
PG 0804+761 & PL+THERM  & 2.08~$\pm$~0.04 & 0.423~$\pm$~0.042 & $\sim$~0 & 249/216\\
 & PL+DISCBB & 2.17~$\pm$~0.03 & 0.168~$\pm$~0.011 & & 258/217\\ 
Mrk 1383 & PL+THERM & 2.12~$\pm$~0.02 & 0.227~$\pm$~0.010 & (9~$\pm$~2)~$\times$~10$^{-3}$ & 282/255 \\
 & PL+DISCBB & 2.11~$\pm$~0.02 & 0.136~$\pm$~0.004 & & 292/256\\
Mrk 876 & PL+THERM & 1.89~$\pm$~0.02 & 0.192~$\pm$~0.006 & 0.017~$\pm$~0.004 & 357/340\\
 & PL+DISCBB & 1.87~$\pm$~0.02 & 0.117~$\pm$~0.003 & & 384/341\\
B2 1028+31 & PL+THERM & 1.72~$\pm$~0.03 & 0.460~$\pm$~0.028 & $\sim$~0 & 496/459\\
 & PL+DISCBB & 1.81~$\pm$~0.02 & 0.167~$\pm$~0.005 & & 505/460\\
B2 1128+31& PL+THERM & 1.80~$\pm$~0.03 & 0.414~$\pm$~0.030 & $\sim$~0 & 376/329\\
 & PL+DISCBB & 1.84~$\pm$~0.02 & 0.157~$\pm$~0.005 & & 390/330\\
B2 1721+34 & PL+THERM & 1.81~$\pm$~0.02 & 0.335~$\pm$~0.029 & $\sim$~0 & 429/358\\
& PL+DISCBB & 1.84~$\pm$~0.01 & 0.143~$\pm$~0.004 & & 435/359\\
\hline
\end{tabular}
\end{center}
\end{table*}

\subsubsection{Comptonization}

The multiple-BB model is a rather na\"{\i}ve way to fit the
soft excess, since the temperatures required  are generally
considerably in excess of the hottest thermal emission expected from
an accretion disc surrounding a 10$^{6}$--10$^{9}~\msun$ black
hole ($\sim$~60~eV for M$_{BH}$~$\sim~$10$^{6}\msun$; see Equation 1, later). 

A more physical approach, as mentioned in the introduction, involves the
Comptonization of disc photons: through this method, the
relatively cool photons from the disc can be up-scattered by hot
distributions of electrons, to form the broad soft excess which is
observed.  The external model {\it thCompfe} (Zdziarski, Johnson \& Magdziarz 1996) was used to fit the
QSO soft excesses, again with an underlying power-law for the higher-energy part of the spectra. As mentioned above, the disc temperatures in these objects are likely to be low, generally $\lo$~60~eV. Subsequently, it will not be easy to fit such a value accurately, because a blackbody at that temperature would be too cool to be visible over the {\it XMM-Newton} band. To avoid this problem, the temperature was estimated by assuming that the bolometric luminosity of the object was 10~$\times$ the 0.5--10~keV value (given in Table~\ref{bb}) and that this is also the Eddington luminosity; this was then used to estimate the mass of the black hole. Accretion was taken to be at the Eddington limit and this, along with the black hole mass, was then substituted into the following equation to obtain an estimate for the temperature of the accretion disc (Peterson 1997):

\begin{equation}
T(r) \sim
6.3\times10^{5}\left(\frac{\dot{M}}{\dot{M}_{Edd}}\right)^{1/4}M_{8}^{-1/4}\left(\frac{r}{R_{sch}}\right)^{-3/4} K
\label{tempeqn}
\end{equation} 

\noindent (where $\dot{M}$ is the mass accretion rate, $\dot{M}_{Edd}$
is the Eddington accretion rate, M$_{8}$ signifies the mass of the central black hole in units of
10$^{8}$ $\msun$ and R$_{sch}$ is the Schwarzschild radius, 2GM/c$^{2}$). The calculated values, rounded to the nearest 5~eV, are listed in the second column of Table~\ref{thcomp}. It should be noted, though, that the resulting parameters of the Comptonization model do not depend strongly on this input temperature for such low values of kT.

The Compton y-parameter gives an indication of the strength of the interaction between the photons and electrons. It is defined as being the average fractional energy change per scattering multiplied by the mean number of interactions and is given (for non-relativistic, optically thick electrons) by

\begin{equation}
y = \frac{4kT}{m_{e}c^{2}}\rm{Max} (\tau,\tau^{2})
\label{y}
\end{equation}

\noindent where $\tau$ is the optical depth ($\tau$ is for the case where the gas is optically thin, $\tau^{2}$ if optically thick), and kT the temperature, of the electron corona; m$_{e}$ is the electronic mass (0.511~MeV/c$^{2}$). The photon index of the spectrum is then given by (Sunyaev \& Titarchuk 1980)

\begin{equation}
\Gamma = \left(\frac{4}{y} + \frac{9}{4}\right)^{1/2} - \frac{1}{2} 
\label{gammaeqn}
\end{equation}

If y~$<<$~1, a modified BB spectrum is observed, with a temperature close to that of the initial input photons. For y~$>>$~1, Comptonization becomes saturated, forming a Wien spectrum, $\propto$~$\nu^{3}$e$^{-h\nu/kT}$, tending towards the temperature of the Comptonizing population of electrons. When y~$\sim$~1, the regime is known as unsaturated Comptonization; this is what is observed here (Table~\ref{thcomp}). In this case, a power-law spectrum is formed over a limited energy range, with an exponential roll-over at $\sim$~4kT. [See Rybicki \& Lightman (1979) for more details.]

\begin{figure*}
\begin{center}
\includegraphics[clip,width=3.8cm,angle=-90]{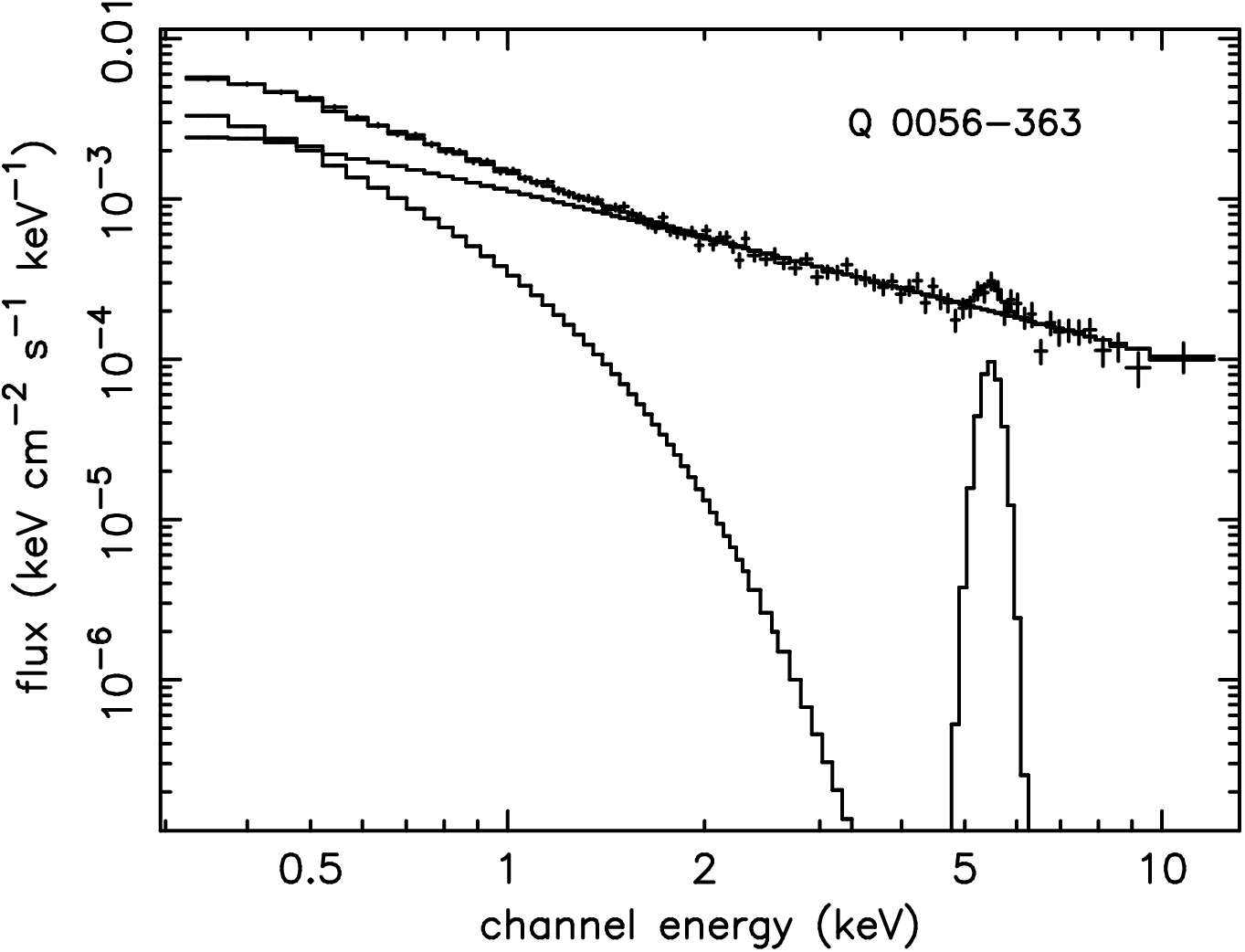}\hspace*{2.0cm}
\includegraphics[clip,width=3.8cm,angle=-90]{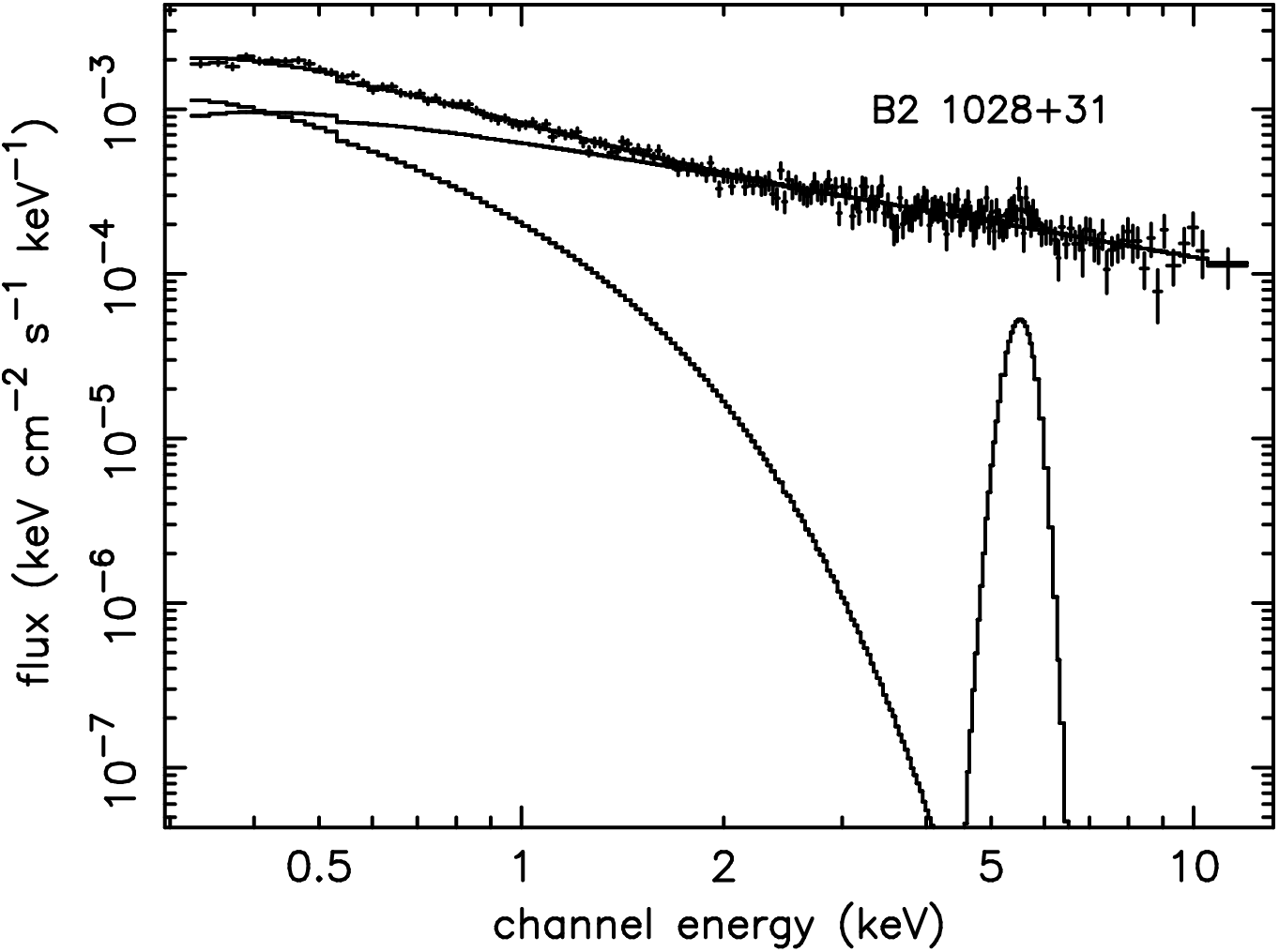}\vspace*{0.005cm}
\includegraphics[clip,width=1.79cm,angle=-90]{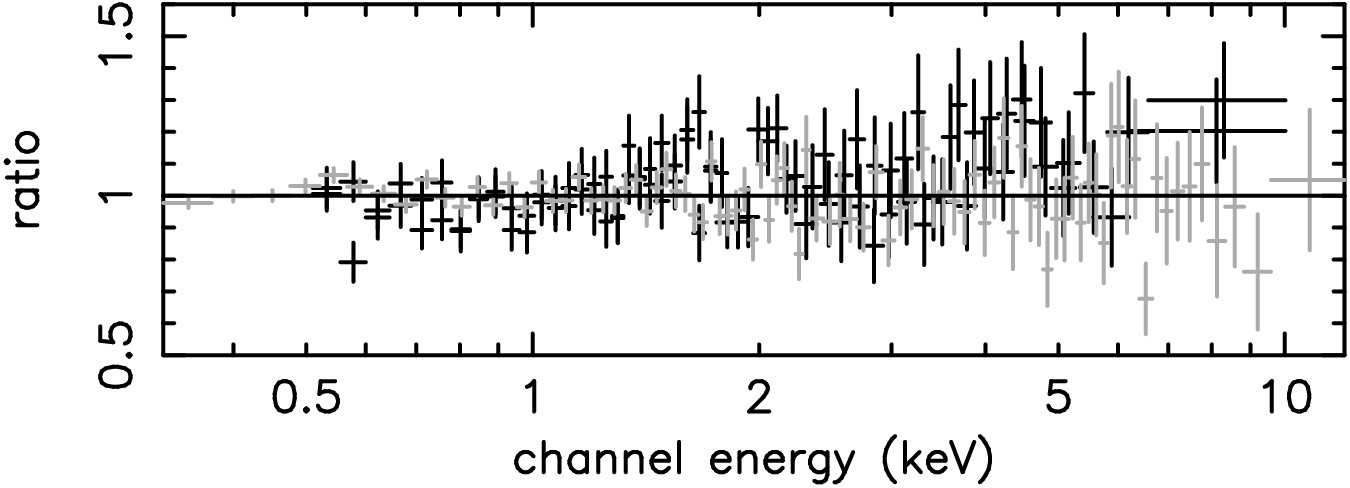}\hspace*{2.0cm}
\includegraphics[clip,width=1.85cm,angle=-90]{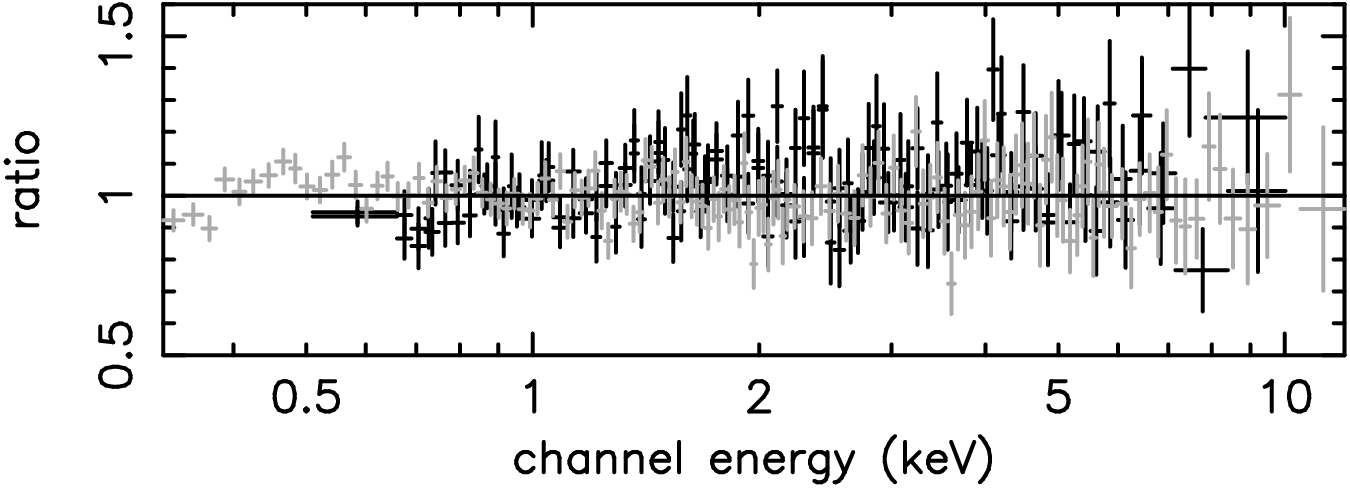}\vspace*{0.1cm}

\includegraphics[clip,width=3.8cm,angle=-90]{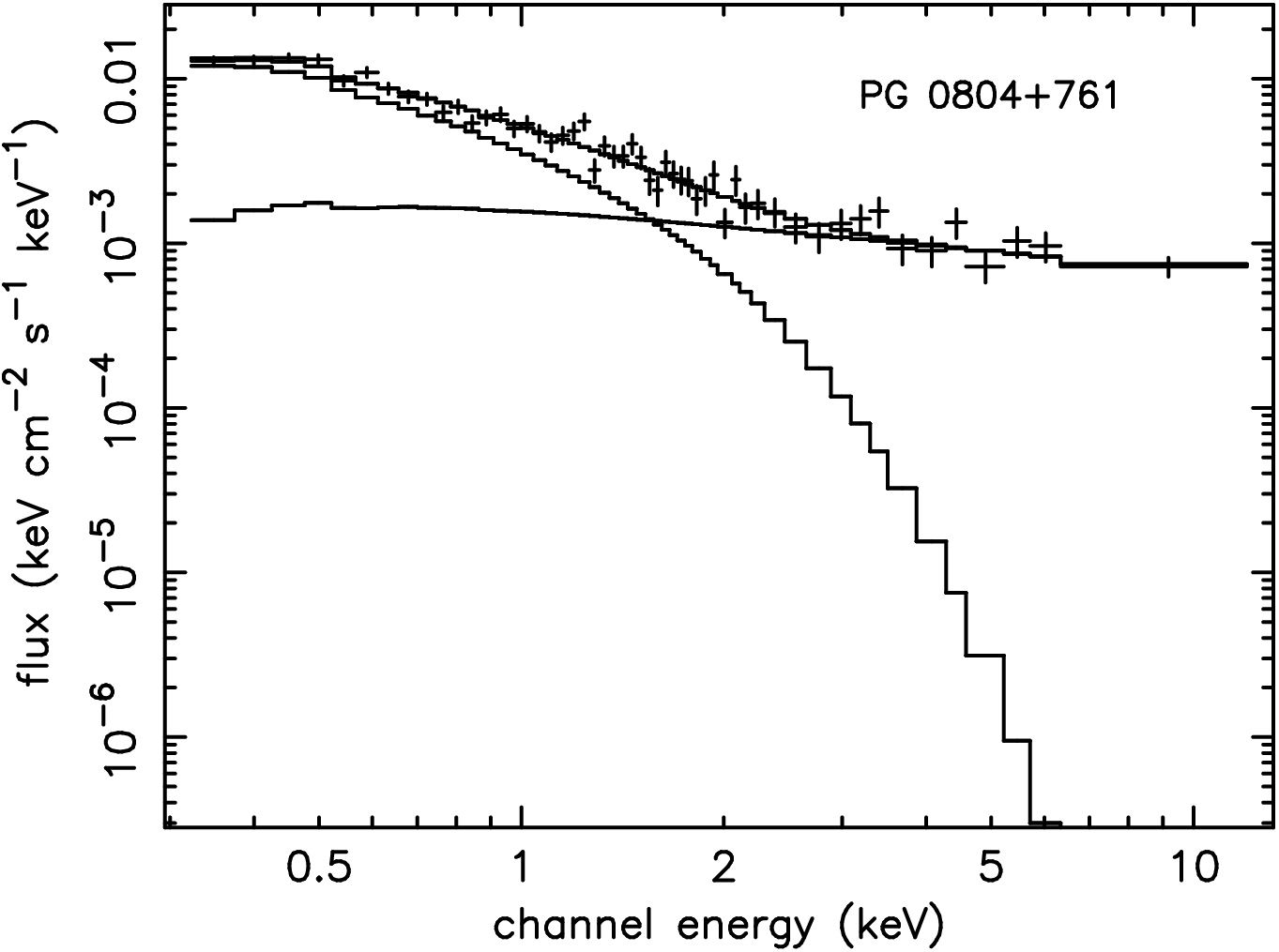}\hspace*{2.0cm}
\includegraphics[clip,width=3.8cm,angle=-90]{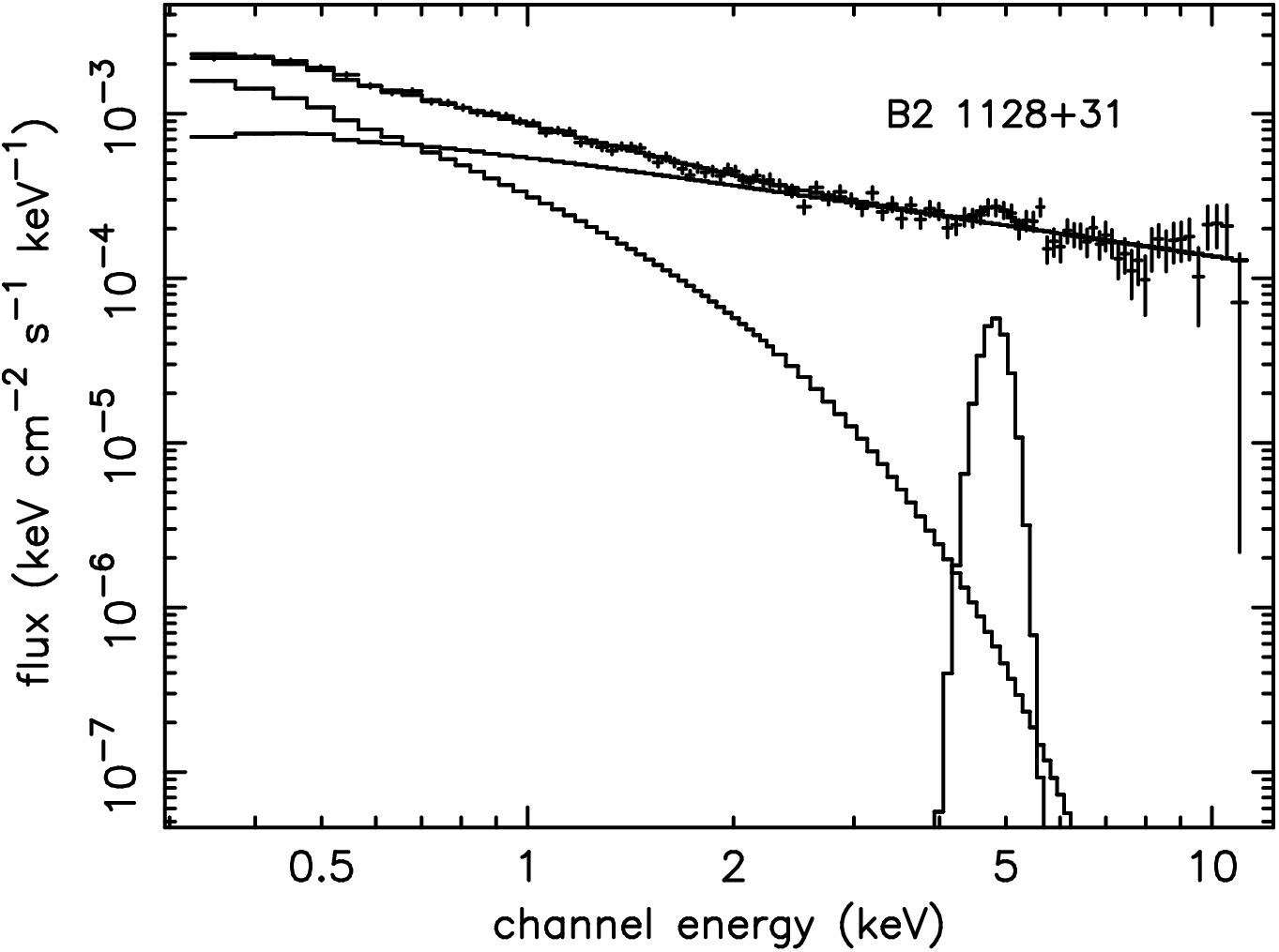}\vspace*{0.005cm}
\includegraphics[clip,width=1.79cm,angle=-90]{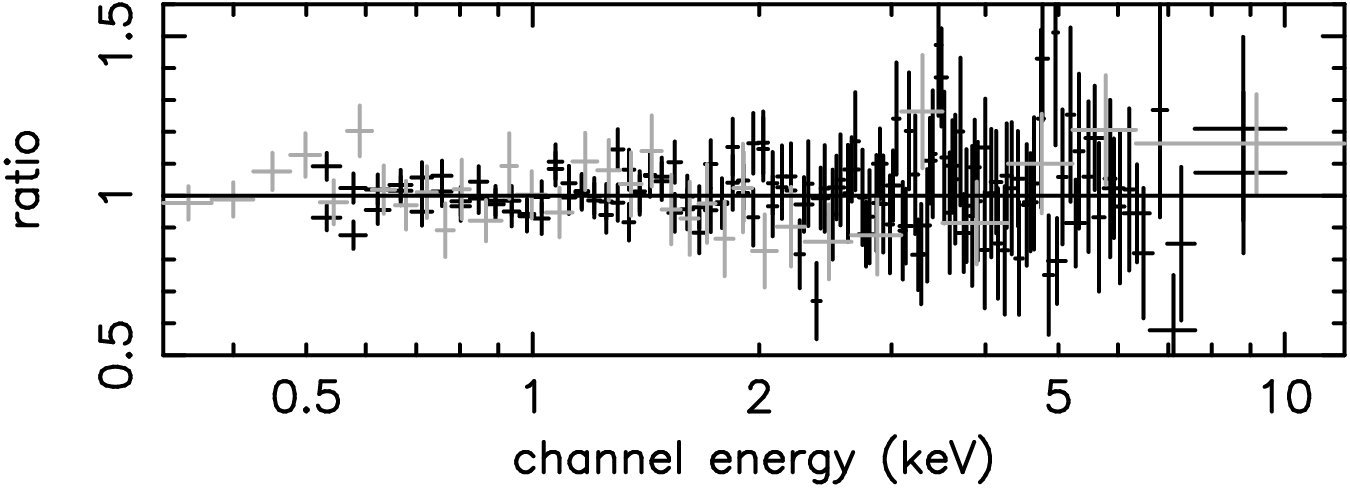}\hspace*{2.0cm}
\includegraphics[clip,width=1.85cm,angle=-90]{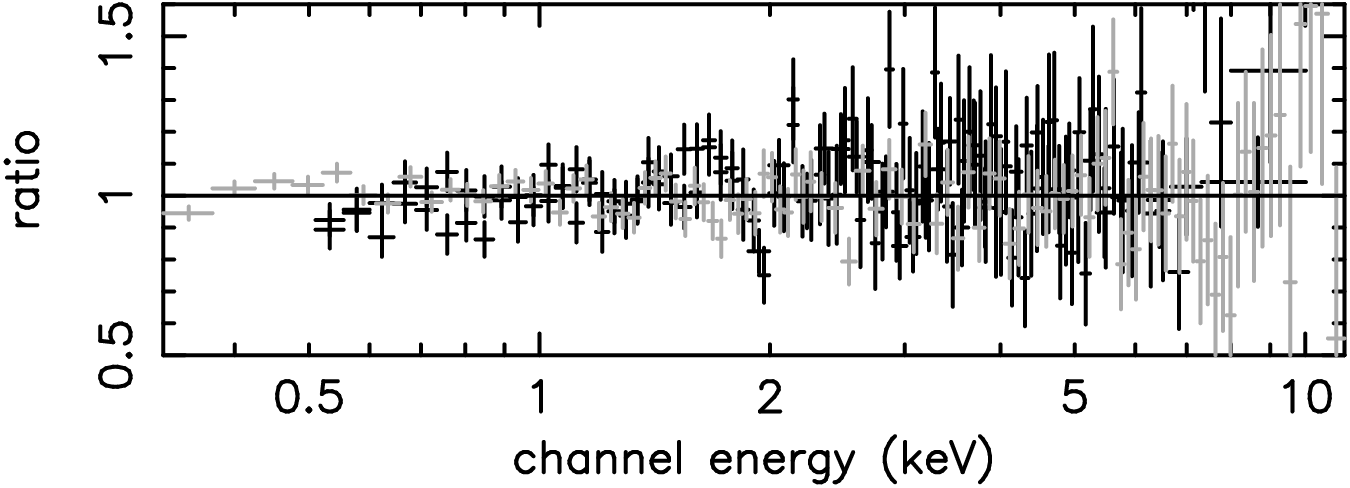}\vspace*{0.1cm}

\includegraphics[clip,width=3.8cm,angle=-90]{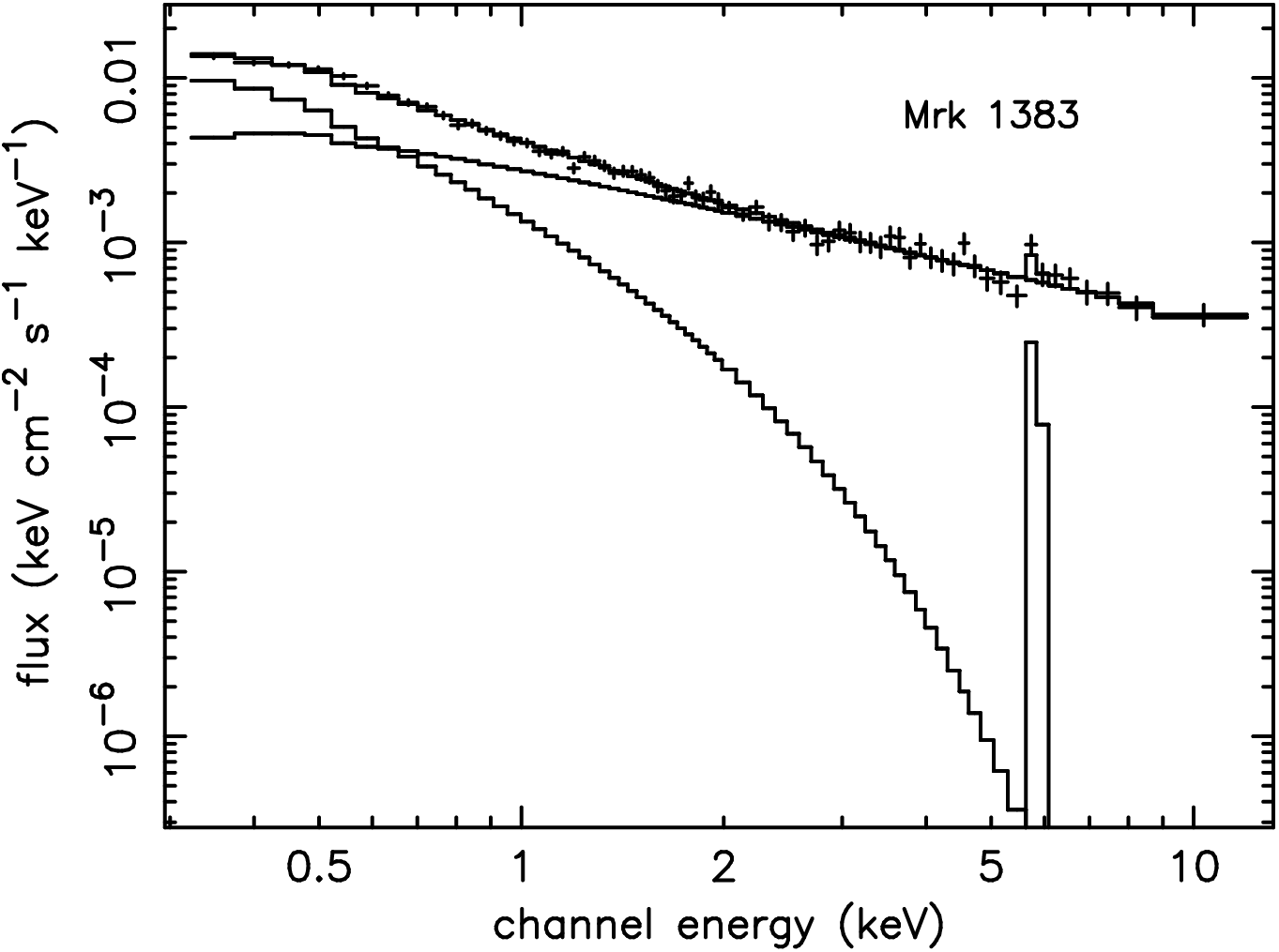}\hspace*{2.0cm}
\includegraphics[clip,width=3.8cm,angle=-90]{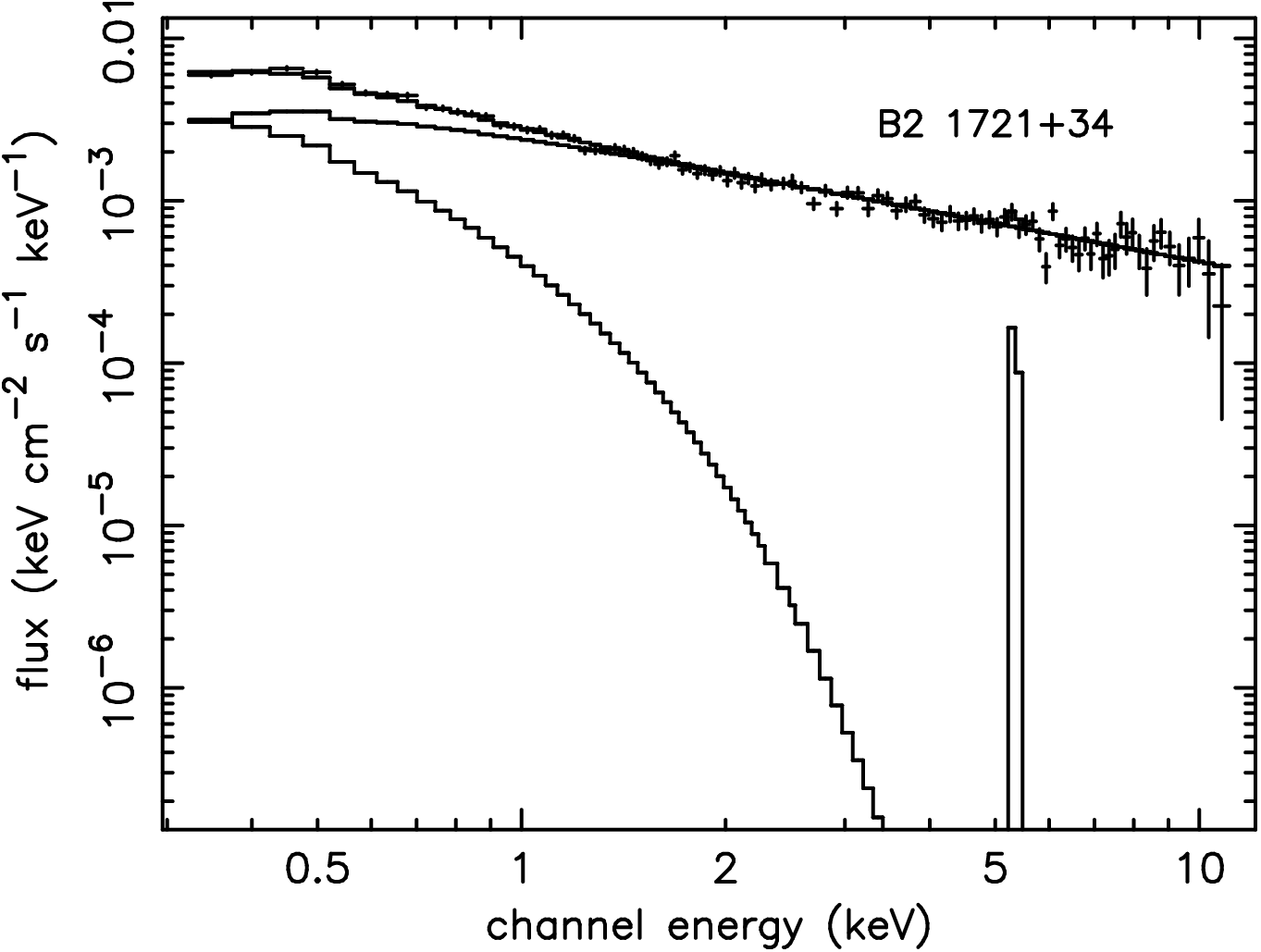}\vspace*{0.005cm}
\includegraphics[clip,width=1.79cm,angle=-90]{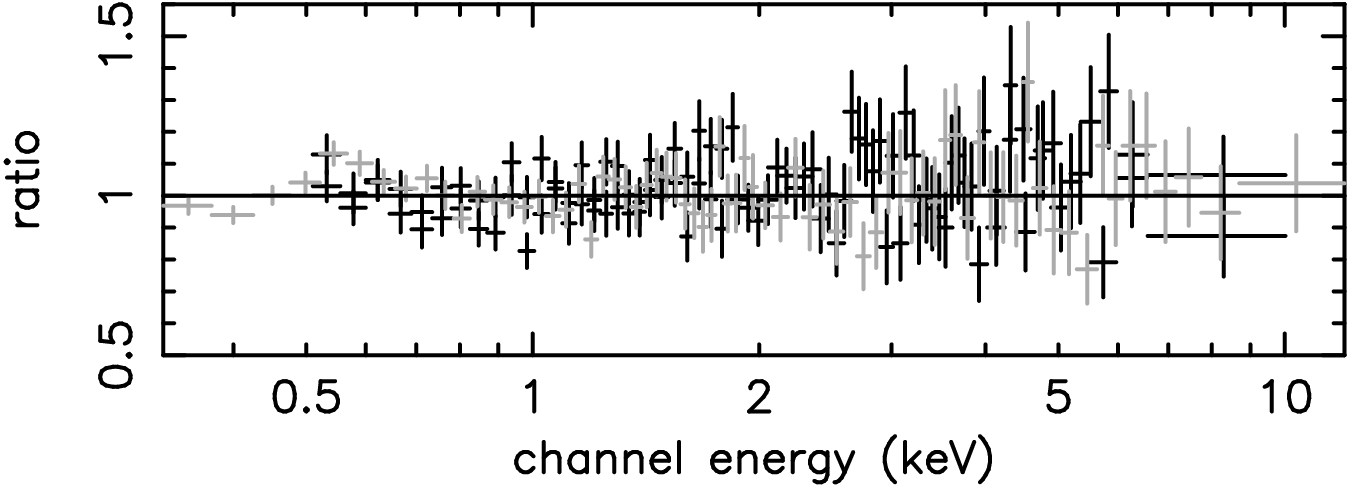}\hspace*{2.0cm}
\includegraphics[clip,width=1.85cm,angle=-90]{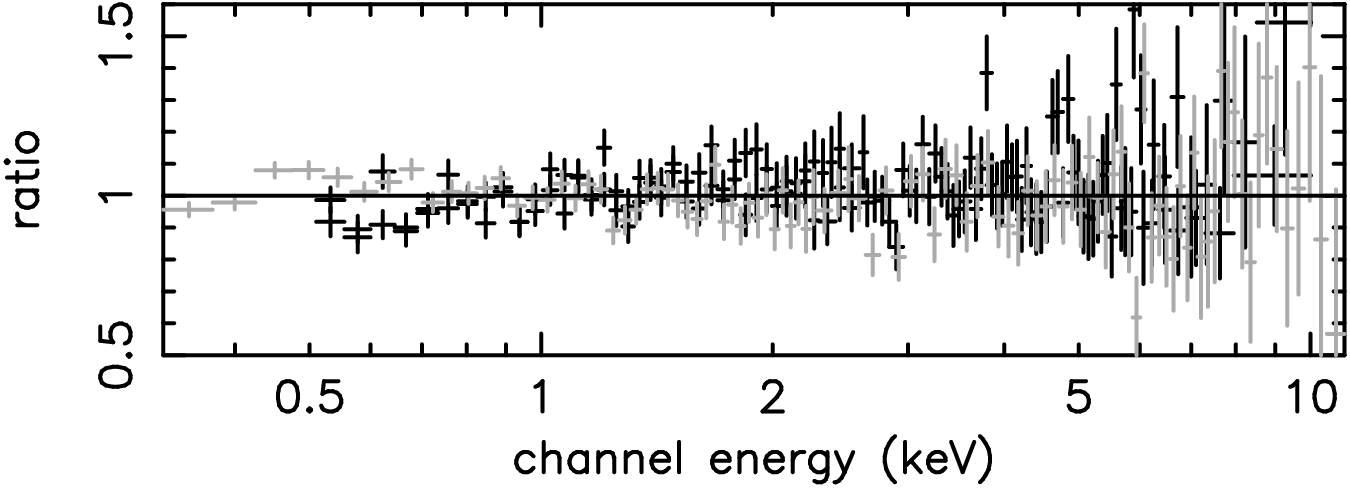}\vspace*{0.1cm}

\includegraphics[clip,width=4.0cm,angle=-90]{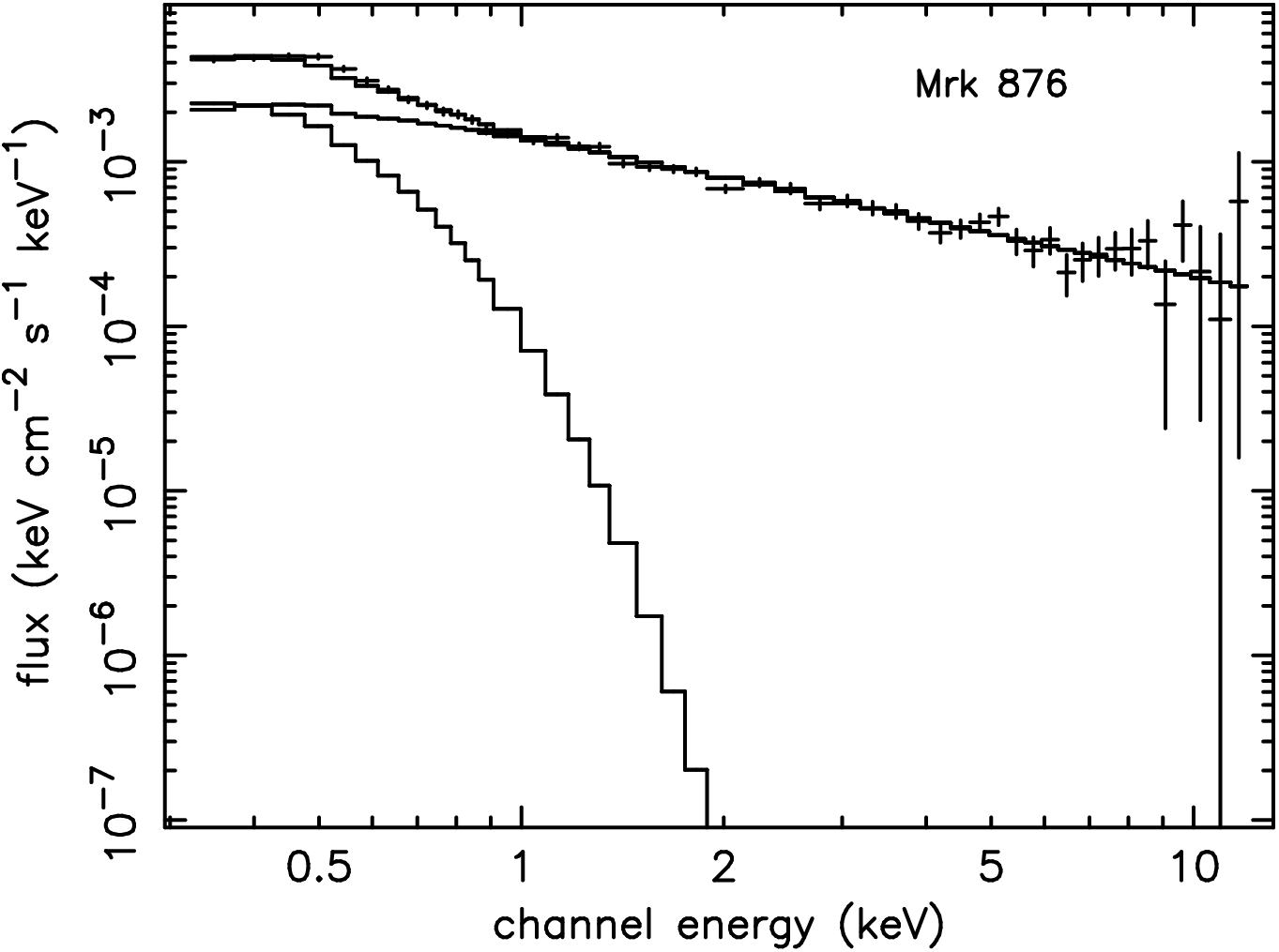}\vspace*{0.01cm}
\hspace{7.0cm}
\includegraphics[clip,width=1.9cm,angle=-90]{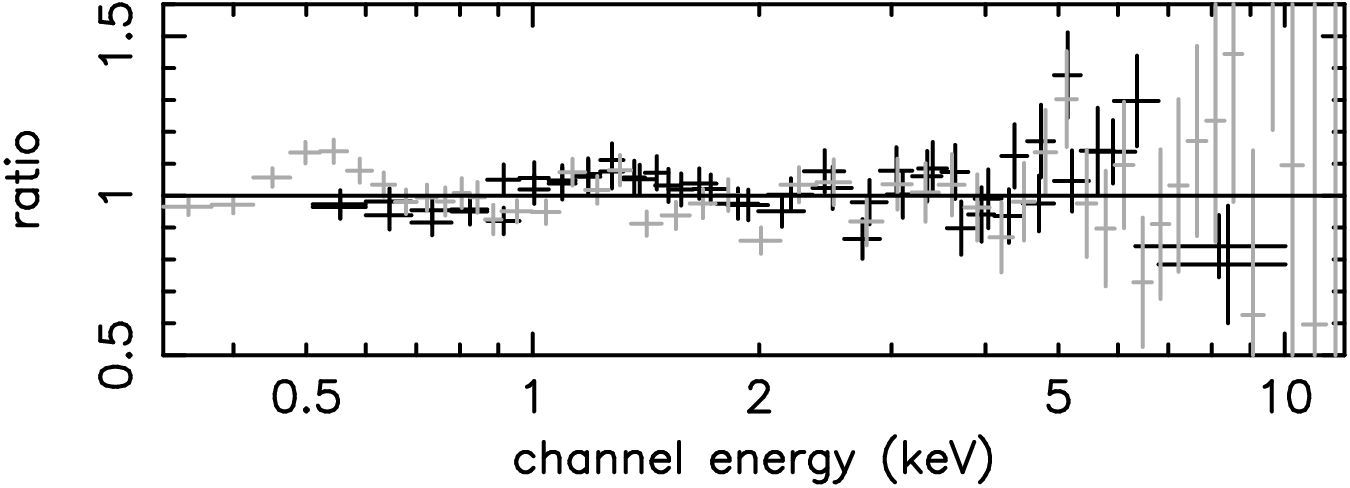}

\caption{Comptonization fits to the soft excess, with an underlying power-law modelling the high-energy spectrum. Only PN data are shown in the unfolded plots, for clarity. In the ratio plots, MOS data are shown in black, PN in grey.}
\label{thcomp_euf}
\end{center}
\end{figure*}

As Table~\ref{thcomp} shows, the Comptonization model lead to statistically good fits,
though generally not quite as good as the simple BB parametrisation; Figure~\ref{thcomp_euf} shows the unfolded fits to the spectra. It is possible that the underlying power-law is also formed through Comptonization, at least in the case of the radio-quiet QSOs. The power-law continua of radio-loud quasars, in particular core-dominated AGN, is related to non-thermal emission from the radio jets, and extends up to several hundred keV, although there may also be a Comptonized component similar to the RQQs. However, to derive any precise constraints on such a Comptonization fit would require a bandpass extending to higher energies than {\it XMM-Newton}.

Because of the close coupling between the temperature and optical depth (Equations~\ref{y} and \ref{gammaeqn}), it is possible that other combinations of values also lead to adequate fits -- for example, a higher electron temperature with a smaller optical depth; such fits have been presented in previous work (e.g., Czerny \& Elvis 1987; Fiore \etal 1995).


It was found that four of the spectra (PG~0804+761, Mrk~1383, B2~1128+31 and B2~1721+34) could be modelled just as well statistically with kT~$\sim$~50~keV (with correspondingly lower optical depths) as with the lower-temperature Comptonised component. This is formally the same as fitting with a second power-law, since the EPIC instruments cannot determine the value of the electron temperature if kT~$\go$~2--3~keV: the exponential roll-over for a Comptonized spectrum occurs at $\go$~4~kT, and, below this point, the emission appears as a power-law. Thus, if 4~kT is greater than $\sim$~10~keV, the temperature cannot be constrained over the {\it XMM-Newton} band, and the fit is indistinguishable from a simple power-law over the same energy range. Because of this, the resultant $\chi^{2}$ values for the high-temperature Comptonization fits to these four spectra are virtually identical to those for the double power-law fits (Table~\ref{pofits}). In Table~\ref{thcomp}, the 1$\sigma$ lower limits are given for the temperature of these components (and the corresponding upper limit for the optical depth), based on the low-kT fit.


Thus, all that can be said for PG~0804+761, Mrk~1383, B2~1128+31 and B2~1721+34, is that the soft excess can be modelled either as low temperature (few 100~eV) Comptonized emission, or as a power-law (which could be formed through high temperature, $\go$ few keV, Comptonization). Q~0056$-$363, Mrk~876 and B2~1028+31 are noticeably better fitted with a lower-temperature component, though.



\begin{table*}
\begin{center}
\caption{Comptonization fits to the soft excess; $\Gamma_{he}$ gives the underlying photon index which models the high energy part of the spectrum. The fits
also include iron lines, with energies and widths fixed at the values
previously found (Table~\ref{hard}). For the QSOs where the soft excess can be modelled as well statistically with a high temperature Comptonization (power-law) component, the lower/upper limits are given for kT/$\tau$; see text for more details.} 
\label{thcomp}
$^{f}$ frozen
$^{a}$ y-parameter, as defined by Equation 2.
\begin{tabular}{p{1.8truecm}p{1.7truecm}p{1.8truecm}p{1.2truecm}p{1.5truecm}p{1.6truecm}p{1.6truecm}p{1.5truecm}p{1.8truecm}}
\hline
 &   \multicolumn{4}{c} {\sc comptonized component}\\
object & input BB & kT  & optical & y$^{a}$   & $\Gamma_{he}$  & $\chi^{2}$/dof \\
 & temp.$^{f}$ (eV) & (keV) & depth   \\ 
\hline
Q~0056$-$363 & 25 & 0.271~$\pm$~0.044 & 15~$\pm$~2 & 0.50~$\pm$~0.04 & 2.71~$\pm$~0.12 & 284/265\\
PG~0804+761 & 25 & $>$ 0.469  & $<$ 12 & 0.58~$\pm$~0.03   & 1.53~$\pm$~0.30 & 230/216\\
Mrk~1383 & 25 & $>$ 0.262 & $<$ 13 & 0.40~$\pm$~0.03 & 1.91~$\pm$~0.10 & 297/255\\
Mrk~876 & 25 & 0.121~$\pm$~0.020 & 38~$\pm$~15  & 1.35~$\pm$~0.72 &  1.87~$\pm$~0.02  & 388/340\\
B2~1028+31 & 25 & 0.333~$\pm$~0.057 & 15~$\pm$~2 & 0.61~$\pm$~0.05   & 1.71~$\pm$~0.04 &  497/459\\
B2~1128+31 & 20  & $>$ 0.424 & $<$ 13  & 0.54~$\pm$~0.03   & 1.62~$\pm$~0.09 &  366/329\\
B2~1721+34 & 20 & $>$ 0.220 & $<$ 18 & 0.50~$\pm$~0.07   & 1.79~$\pm$~0.03 & 430/358\\
\hline
\end{tabular}
\end{center}
\end{table*}

\subsubsection{B2 1028+31}

The RLQ B2~1028+31 is located in the centre of the weak Abell cluster A1030. Because of this, there is a possibility that the X-ray emission from the AGN could be contaminated by the intra-cluster gas.

The temperature of the extended emission is $\sim$~0.2~keV (Sarazin \etal  1999), which is close to the hotter of the blackbodies used here to model the AGN soft excess; this is much cooler than expected for the intra-cluster medium (5--10~keV). Sarazin \etal (1999) find that the X-ray emission is dominated by the quasar, with the extended soft emission possibly due to inverse Compton emission from the radio lobes. They state that X-rays from the intra-cluster medium itself make an insignificant contribution to the X-ray luminosity of $\lo$~8~per~cent.

Such extended emission has been frequently detected in other RLQs and radio galaxies (see, for example, Scharf \etal 2003; Fabian, Celotti \& Johnstone 2003; Comastri \etal 2003; Siemiginowska \etal 2003; Chartas \etal 2000), so may also be relevant for B2~1128+31 and B2~1721+34 as well. However, it is unlikely that this emission would significantly affect the results presented here, given that the extraction region is only 40~arcsec in radius.

\subsection{Ionized disc models}
\label{reflection}

Some of the objects in this sample show iron lines with finite widths; this indicates that the lines may be formed towards the inner accretion disc. In order to investigate this possibility, the observed frame 2--12~keV spectra were fitted with the ionized disc reflection model described by Ballantyne \etal (2001). It was found that all seven spectra, even those which were better fitted by a {\it narrow} Gaussian component, could be well described by this reflection model. The results of the fits are given in Table~\ref{ref}. It is not surprising that the ionization parameters found are low, since, when fitting Gaussian lines, the energies were close to 6.4~keV, signifying neutral iron. The line in the spectrum of PG~0804+761, however, was found to be slightly ionized (E~$\sim$~6.7~keV) and is modelled here by the highest ionization parameter.

\begin{table}
\begin{center}
\caption{The fits to the 2--10~keV spectra using the disc reflection model of Ballantyne \etal (2001). $^{a}$ ionization parameter -- $\xi$~=~4$\pi$F$_{x}$/N$_{H}$; $^{b}$ reflection component -- R~=~$\Omega$/2$\pi$.} 
\label{ref}
\begin{tabular}{p{1.6truecm}p{1.3truecm}p{1.2truecm}p{1.2truecm}p{1.2truecm}p{0.9truecm}}
\hline

object  &  $\xi^{a}$  & $\Gamma$ & R$^{b}$ & $\chi^{2}$/dof\\
& (erg~cm~s$^{-1}$)\\
\hline
Q~0056$-$363 & 1.64~$\pm$~0.21 & 2.06~$\pm$~0.05 & 1.10~$\pm$~0.44 & 143/169\\
PG~0804+761 & 3.06~$\pm$~0.33 & 1.98~$\pm$~0.07 & 0.32~$\pm$~0.27 & 106/120\\ 
Mrk~1383 & 1.47~$\pm$~0.35 & 2.01~$\pm$~0.05 & 0.64~$\pm$~0.36 & 157/158\\
Mrk~876 & 2.25~$\pm$~0.31 & 1.88~$\pm$~0.06 & 0.77~$\pm$~0.30 & 210/240\\
B2~1028+31& 1.92~$\pm$~0.13 & 1.72~$\pm$~0.03 & 0.96~$\pm$~0.38 & 248/278\\
B2~1128+31 & 2.04~$\pm$~0.12 & 1.76~$\pm$~0.04 & 1.14~$\pm$~0.38 & 229/233\\
B2~1721+34 & 1.71~$\pm$~0.21 & 1.79~$\pm$~0.03 & 0.41~$\pm$~0.25 & 235/261\\ 
\hline
\end{tabular}
\end{center}
\end{table}

Ionized reflection may also contribute to the soft emission observed in broad-band X-ray spectra. Sometimes the reflection can account for the entire soft excess (e.g., Mrk~205, Reeves \etal 2001), while, in other spectra, the strength of the excess is simply decreased, shown by either the requirement for fewer blackbodies (e.g., Mrk~359, O'Brien \etal 2001), or by a lower normalisation of the components (e.g., Mrk~896, Page \etal 2003). For each of the QSOs in this sample, it is this last case -- the decreased normalisation -- which is observed.



\section{Discussion}
\label{sec:disc}
\subsection{Iron lines}

Observations with {\it Ginga} found that many AGN spectra showed evidence for an emission line around 6.4~keV, corresponding to the Fe~K$\alpha$ line (Pounds \etal 1989, 1990; Nandra \etal 1991; Nandra \& Pounds 1994). In general, {\it XMM-Newton} observations find the lines to be unresolved by the instrument ($\sigma$~$<$~10~eV), with relatively few broad lines. Examples of resolved line widths include MCG~$-$6$-$30$-$15 (Fabian \etal 2002), Mrk~205 (Reeves \etal 2001) and Mrk~509 (Pounds \etal 2001).

No iron lines have been reported in any of these objects from previous X-ray observations, although the {\it XMM-Newton} spectrum of Q~0056$-$363 has been recently published by Porquet \& Reeves (2003). They identified a strong broad Fe~K$\alpha$ line and the results presented in this paper are in complete agreement with their findings. Lawson \& Turner (1997) analysed {\it Ginga} spectra of PG~0804+761, Mrk~1383, Mrk~876 and B2~1721+34, but found that iron emission lines were not significant. PG~0804+761 was also observed with {\it ASCA} (George \etal 2000) where, again, there was no evidence for an iron line. In the {\it XMM-Newton} spectra presented here, statistically significant narrow lines were, however, found in Mrk~1383 and B2~1721+34, as well as resolved lines in Q~0056$-$363, B2~1028+31 and B2~1128+31.

If the emission lines in Q~0056$-$363, B2~1028+31 and B2~1128+31 presented here are believed to be broadened, then the velocity widths are found to be $\sim$~(1--1.5)~$\times$~10$^{4}$ km~s$^{-1}$. This indicates that the emission originates in the inner accretion disc, close to the black hole, rather than in distant matter, such as the molecular torus. However, it must be noted that the lines are consistent with being intrinsically narrow at the 99~per~cent level.

\subsection{Soft excess}

Each of the seven QSOs in this sample have been found to show soft excess emission, as is commonly the case for AGN. 

Although Q~0056$-$363 has only been previously observed with {\it ROSAT}, {\it ASCA} (George \etal 2000) and {\it EXOSAT} (Saxton \etal 1993) observations of PG~0804+761 have also occurred, both of which found excess soft emission. 

Likewise, {\it EXOSAT} (Comastri \etal 1992) identified a soft excess in Mrk~1383 (which they found to be better modelled as a blackbody, rather than a second power-law, agreeing with the data in this paper), while Wilkes \& Elvis (1987) found that their {\it Einstein} spectrum required less than Galactic absorption -- often an indication of soft excess emission. {\it Einstein} spectra of Mrk~876 (Masnou \etal 1992; Wilkes \& Elvis 1987) did not reveal a soft excess, though.
 
B2~1028+31 has been previously found to possess a soft excess by {\it Einstein} (Wilkes \& Elvis 1987; Masnou \etal 1992), {\it ROSAT} and {\it ASCA} (Sarazin \etal 1999), with the excess being fitted by either a blackbody, or with a broken power-law (over the whole energy band). Sarazin \etal (1999) found the soft excess to contribute $\sim$~15~per~cent of the total X-ray luminosity of the quasar, in complete agreement with the {\it XMM-Newton} data presented here. B2~1128+31 and B2~1721+34 do not appear to have had soft excess reported prior to this paper. B2~1721+34 is, however, the largest radio source associated with a quasar (Barthel \etal 1989) and shows superluminal motion.

Although modelling the soft excess in individual objects provides some useful information, comparing a number of objects may lead to a more detailed understanding of the processes involved. Of the different methods for modelling the soft excess covered in this paper (blackbodies, thermal plasma, disc blackbodies and Comptonization), the multiple blackbody fit is statistically the best for all but Mrk~876 (which is better fitted by the thermal {\it mekal} model). The disc blackbody results in the worst $\chi^{2}$ values. Although the soft excess has been thoroughly investigated since it was first discovered in 1985, the actual form of the emission is still uncertain. The multiple blackbody model generally gives the best results, but it is difficult to see how such emission could physically come about; Comptonization can easily explain the high temperatures observed, though. However, it is still not possible to state with any certainty what the origin of the soft excess truly is.



\begin{figure}
\begin{center}
\includegraphics[clip,width=6.0cm,angle=-90]{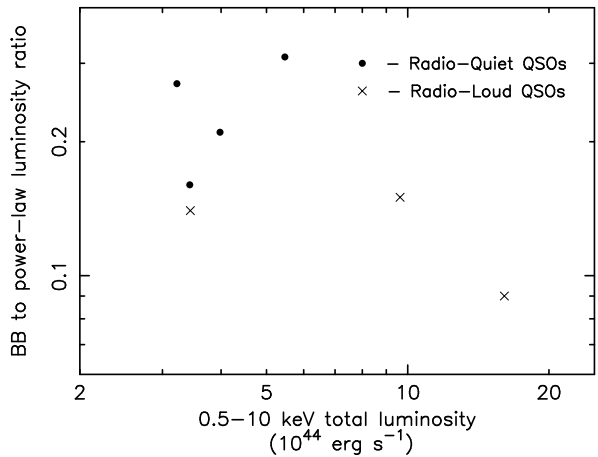}
\caption{The strength of the soft excess does not seem to depend on the luminosity of the object.}
\label{strength-lum}
\end{center}
\end{figure}
\begin{figure}
\begin{center}
\includegraphics[clip,width=6.0cm,angle=-90]{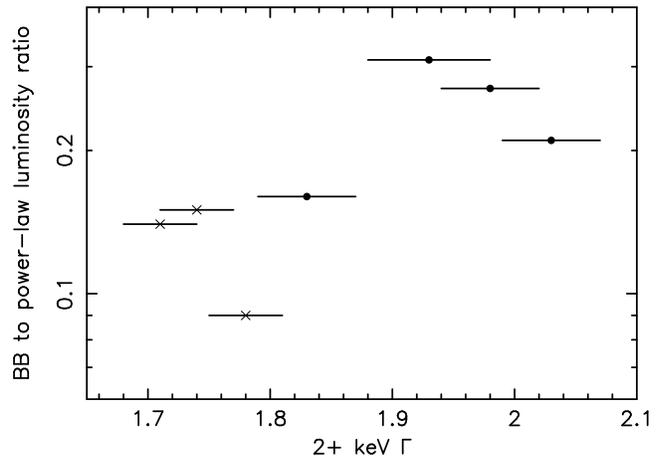}
\caption{Plotting the strength of the soft excess against the 2--10~keV power-law slope reveals a possible weak correlation. Symbols as in Figure~\ref{strength-lum}.}
\label{strength-gamma}
\end{center}
\end{figure}

In order to try and identify any existing trends, various figures were plotted and are discussed below. The relative strengths of the blackbody components compared to the power-laws were given in Table~\ref{bb}. In Figure~\ref{strength-lum}, this ratio is plotted against the broad-band 0.5--10~keV luminosity for each object. There is no correlation between the values (probability of only 68~per~cent for a negative correlation, using Spearman Rank analysis). Figure~\ref{strength-gamma} shows that there may be a weak relationship between the 2--10~keV photon index and the relative strength of the soft excess for these QSOs ($\sim$~96~per~cent for a positive correlation between the values). It has been previously found that Narrow Line Seyfert 1 galaxies show both steep 2--10~keV power-law slopes (Brandt, Mathur \& Elvis 1997) and strong soft excesses (corresponding to steep spectra over the {\it ROSAT} band; Boller, Brandt \& Fink 1996); this could be due to a Compton cooling effect, whereby the excess soft flux cools the power-law electrons. The slight correlation between slope and soft excess found here could also be related to this effect. 

\begin{figure}
\begin{center}
\includegraphics[clip,width=6.0cm,angle=-90]{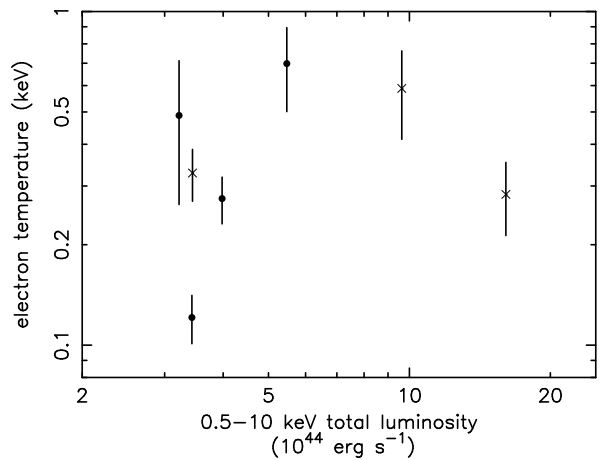}
\caption{There is no apparent relationship between the temperature of the soft excess and the 0.5--10~keV broad-band luminosity. Symbols as in Figure~\ref{strength-lum}.}
\label{kt-lum}
\end{center}
\end{figure}

Figure~\ref{kt-lum} plots the Comptonized temperature of the soft excess against the broad-band object luminosity. There is, again, no trend between the values (probability of only 42~per~cent from Spearman Rank).

From the three previous plots it is noticeable that there is little difference in the soft excess properties of the RQQs and RLQs, with the temperatures covering a very similar range. The radio-quiet objects show slightly stronger soft excess luminosity ratios, but the values are not widely spread; there is only a factor of three between the weakest and strongest component. The mean luminosity ratio is 0.24 for the RQQs and 0.13 for the RLQs. Using Student's T-test, there is a probability of 0.046 that the difference is due to chance and that the two samples come from the same group. In other words, there is a $\sim$~95~per~cent probability that the RQQs and RLQs are separate populations when considering the strength of the soft excess. This indicates that there may be a real difference between the groups, but is, by no means, conclusive. This (possible) difference in populations could explain the weak correlation between the soft excess luminosity ratio and the power-law indices above 2~keV: RLQs have flatter slopes and (possibly) weaker soft excesses, with RQQs showing steeper photon indices and more luminous soft excesses. One reason for RLQs having weaker soft excesses could be the dilution of the luminosity ratio by the non-thermal X-ray emission from the radio-jets.

Although this sample is not large, these results indicate that the soft excesses found in radio-quiet and radio-loud AGN are similar, although radio-loud quasars do possess flatter high-energy photon indices. This flatter slope is likely to be due to emission from the relativistic radio jet, implying that the soft excess and radio emission may be unrelated. Also the properties of the soft excess (i.e., `strength' and temperature) do not appear to be dependent on the X-ray luminosity of the source. This is in agreement with Page \etal (2004), who investigated a sample of high luminosity radio-quiet QSOs, finding that the X-ray continuum shape (including the soft excess) remains essentially constant over a wide range of black hole mass and luminosity. Thus, the implication is that the soft excess may be independent of luminosity, black hole mass and radio-loudness. Clearly these results should be investigated for a much larger group of active galaxies, covering a good range of both RQQs and RLQs.

\section{Conclusions}

The soft excesses of four radio-quiet and three radio-loud QSOs are analysed. As a simple parametrization, two blackbody components fit the soft excess very well. More physically, Comptonization of the disc photons is invoked to explain the soft X-ray emission; this model also provides good fits. There is little obvious intrinsic difference between the soft excess in the radio-quiet and radio-loud objects, with the electron temperatures covering the same range. Five of the objects also showed evidence for iron emission, with three of them being better fitted with a somewhat broadened emission line. These widths are not very large, however, and are significant only at the 95~per~cent level.

\section{ACKNOWLEDGMENTS}
The work in this paper is based on observations with {\it
XMM-Newton}, an ESA
science mission, with instruments and contributions directly funded by
ESA and NASA. The authors would like to thank the EPIC Consortium for all their work during the calibration phase, 
and the SOC and SSC teams for making the observation and analysis
possible; also the anonymous referee, whose careful reading and detailed comments improved the paper.
This research has made use of the NASA/IPAC Extragalactic
Database (NED), which is operated by the Jet Propulsion Laboratory,
California Institute of Technology, under contract with the National
Aeronautics and Space Administation.



\end{document}